\documentstyle[12pt,aasms4,psfig]{article}
\newcommand{\etal}{{\it et al.} }
\newcommand{\asca}{{\it ASCA} }
\newcommand{\einstein} {{\it Einstein} } 

\newcommand{\ginga}{{\it Ginga} }

\newcommand{\rosat}{{\it ROSAT} }
\newcommand{\iue}{{\it IUE} }
\newcommand{\fuse}{{\it FUSE} }
\newcommand{\xmm}{{\it XMM-Newton} }
\newcommand{\chandra}{{\it Chandra} }

\newcommand{\hst}{{\it HST}/STIS }   
\newcommand{\rxte}{{\it RXTE} }
\newcommand{\hetg}{{\it HETGS} }
\newcommand{\fekalfa}{{Fe~K$\alpha$} }
\newcommand{\bsax}{{\it BeppoSAX} }
        
\newcommand{\cfour}{C~{\sc iv} }
\newcommand{\csix}{C~{\sc vi} }
\newcommand{\nfive}{N~{\sc v} }
\newcommand{\sifour}{Si~{\sc iv} }

\newcommand{\sila}{Si~{\sc xiv}~Ly$\alpha$ ($\lambda 6.182\AA$) }
\newcommand{\nela}{Ne~{\sc x}~Ly$\alpha$ ($\lambda 12.134\AA$) }
\newcommand{\nila}{N~{\sc vii}~Ly$\alpha$ ($\lambda 24.781\AA$) }
 
\newcommand{\oxla}{O~{\sc viii}~Ly$\alpha$ ($\lambda 18.969\AA$) }
\newcommand{\mgla}{Mg~{\sc xii}~Ly$\alpha$ ($\lambda 8.421\AA$) }
\newcommand{\arla}{Ar~{\sc xviii}~Ly$\alpha$ ($\lambda 3.733\AA$) }
\newcommand{\nala}{Na~{\sc xi}~Ly$\alpha$ ($\lambda 10.025\AA$) }

\newcommand{\nilb}{N~{\sc vii}~Ly$\beta$ ($\lambda 20.910\AA$) }

\newcommand{\oxyseven}{O~{\sc vii} }
\newcommand{\oxyeight}{O~{\sc viii} }

\newcommand{\nenine}{Ne~{\sc ix} }

\newcommand{\neniner}{Ne~{\sc ix} (r) }

\newcommand{\fefourteen}{Fe~{\sc xiv} }
\newcommand{\fetwelve}{Fe~{\sc xii} }
\newcommand{\fefifteen}{Fe~{\sc xv} }
\newcommand{\feseventeen}{Fe~{\sc xvii} }
\newcommand{\feeighteen}{Fe~{\sc xviii} }  
\newcommand{\fenineteen}{Fe~{\sc xix} }

\newcommand{\fetwentyone}{Fe~{\sc xxi} }

\newcommand{\mgelevenr}{Mg~{\sc xi} (r) }

\newcommand{\resonetwo} {${\it  1s^{2}-1s 2p}$ }
\newcommand{\resonethree} {${\it 1s^{2}-1s 3p}$ }
\newcommand{\figufspecnhgal}{Fig.~1 }
\newcommand{\figufspecedges}{Fig.~2 }
\newcommand{\figufspeccont}{Fig.~3 }
\newcommand{\figmegsnr}{Fig.~4 }
\newcommand{\fighegsnr}{Fig.~5 }
\newcommand{\figmeglyman}{Fig.~6 }
\newcommand{\figmegbalmer}{Fig.~7 }
\newcommand{\figvelprof}{Fig.~8 }
\newcommand{\figvelprofone}{Fig.~8 }
\newcommand{\figvelproftwo}{Fig.~9 }
\newcommand{\figcog}{Fig.~10 }
\newcommand{\figsed}{Fig.~11 }
\newcommand{\figxstaroverdata}{Fig.~12 }
\newcommand{\figxstarmodel}{Fig.~13 }
\newcommand{\figxstarcontour}{Fig.~14 }
\newcommand{\figuvvelocities}{Fig.~9 }
\newcommand{\src}{Mrk~509 }
\newcommand{\mk}{Mrk~509 }
\newcommand{\mcg}{MCG~$-$6$-$30$-$15 } 

\begin{document}

\title{THE KINEMATICS AND PHYSICAL CONDITIONS OF THE IONIZED GAS IN MARKARIAN 509. I. CHANDRA HIGH-ENERGY GRATING SPECTROSCOPY}

\author{Tahir Yaqoob\altaffilmark{1,2},
Barry McKernan\altaffilmark{1},
Steven B. Kraemer\altaffilmark{3,4}, D. Michael Crenshaw\altaffilmark{5},
Jack R. Gabel\altaffilmark{3,4},
Ian M. George\altaffilmark{2,6},
T. Jane Turner\altaffilmark{2,6}}

\begin{center}
{\it Accepted for Publication in the Astrophysical Journal, 22 August 2002}
\end{center}

\altaffiltext{1}{Department of Physics and Astronomy,
Johns Hopkins University, Baltimore, MD 21218}
\altaffiltext{2}{Laboratory for High Energy Astrophysics,
NASA/Goddard Space Flight Center, Greenbelt, MD 20771}
\altaffiltext{3}{Department of Physics,
Catholic University of America, 200 Hannan Hall, Washington, DC 20064}
\altaffiltext{4}{Laboratory for Astronomy and Solar Physics, Code 681,
NASA/Goddard Space Flight Center, Greenbelt, MD 20771}
\altaffiltext{5}{Department of Physics and Astronomy, Georgia State University,
Astronomy Offices, One Park Place
South SE, Suite 700, Atlanta, GA 30303}
\altaffiltext{6}{Joint Center for Astrophysics, University of Maryland,
Baltimore County, 1000 Hilltop Circle, Baltimore, MD 21250}

\begin{abstract}

We observed the Seyfert~1 galaxy Mrk~509 for $\sim 59$~ks
with the {\it Chandra} High-Energy Transmission Gratings,
simultaneously with {\it HST}/STIS and {\it RXTE}.
Here we present a detailed analysis of the soft
X-ray spectrum observed with {\it Chandra}. We measure strong absorption
lines from He-like Ne and Mg, and from H-like N, O, and Ne.
Weaker absorption lines may also be present. 
The lines are
unresolved except for \nela and
\nenine \resonetwo ($\lambda 13.447\AA$), which appear to be marginally
resolved. The profiles are blueshifted with respect to the
systemic velocity of Mrk~509, indicating an outflow of
$\sim -200$~$\rm km \ s^{-1}$. There is also a hint that the
profiles may have a velocity component near systemic.
The soft X-ray spectrum can be described in remarkable
detail with a simple, single-zone photoionized absorber
having an equivalent neutral Hydrogen column density of
$2.06^{+0.39}_{-0.45} \rm \ \times 10^{21} \ \rm cm^{-2}$
and an ionization parameter of $\log{\xi} = 1.76^{+0.13}_{-0.14}$
(or $\log{U}=0.27$). Although the photoionized gas almost certainly
is comprised of matter in more than one ionization state 
and may consist of several kinematic components, data with
better spectral resolution and signal-to-noise would be 
required to justify a more complex model.
The UV data, on the other hand, have a velocity resolution
of $\sim 10$~$\rm km \ s^{-1}$ and can easily detect 
eight kinematic components, covering roughly the
same velocities as the X-ray absorption profiles. 
Even though the X-ray and UV absorbers share the same
velocity space, the UV absorbers have a much smaller column
density and ionization state. We show that models of
the X-ray data do not predict significant UV absorption
and are therefore consistent with the UV data.
Finally, we do not detect any soft X-ray emission lines.
\end{abstract}

\keywords{galaxies: active --
galaxies: individual (\src) -- galaxies: Seyfert --
techniques: spectroscopic -- ultraviolet: galaxies 
-- X-rays: galaxies -- X-rays: galaxies}

\section{Introduction}
\label{sec:intro}
 
X-ray and UV absorption and emission by photoionized circumnuclear gas in type~1 Seyfert galaxies
is a key observational diagnostic. While it has been possible
to study the UV absorption with a velocity resolution of
$\sim 10$~$\rm km \ s^{-1}$, the kinematics of the
X-ray absorber observed with CCDs (such as those aboard \asca)
could only be studied with a velocity resolution $>10,000$~$\rm km \ s^{-1}$.
Inadequate spectral resolution, combined with a lack of
simultaneity between X-ray and UV observations
has resulted
in major uncertainties in the dynamics, physical state, location, and
geometry of the X-ray and UV absorbers, as well as the relation between
the two.
The launch of \chandra and \xmm began a new era in the
study of X-ray photoionized circumnuclear gas. The
energy resolution of the \chandra transmission gratings is
currently the best available in the 0.5--10 keV band, and is as high as
$\sim 280$~$\rm km \ s^{-1}$ at 0.5 keV.
High resolution X-ray
spectroscopy with \chandra now allows the gas kinematics to be
studied seriously for the first time, and the detection
of individual absorption and emission lines can now place
very strong constraints on the ionization
structure of the gas.
 
The existence of warm, or partially ionized, X-ray absorbing gas in
type~1 active galactic nuclei (AGNs)
was first suggested by \einstein observations of QSO MR~2251$+$178
(\cite{halpern84}). Subsequently,  \rosat 
and \asca showed this to be a common phenomenon, present
in roughly $\sim 50-60\%$ of
type~1 AGNs, and studies focused on the measurements
of the \oxyseven and \oxyeight
absorption edges which appeared
to be the most prominent features in these low to
moderate resolution spectra (\cite{nandra92}; \cite{fabian94}; \cite{reynolds97}; \cite{george98}).
High-resolution grating observations
with \chandra have since been used to study the X-ray warm absorbers in
several Seyfert~1 galaxies 
(\cite{coll01}; \cite{lee01}; \cite{sako01a}; \cite{pounds01}; \cite{kaastra02}; 
\cite{kaspi02}; \cite{yaqoob02a}).
Discrete, narrow absorption features
(FWHM less than $\sim 2000 \ \rm km \ s^{-1}$),
often unresolved, are found to be typically blueshifted,
with outflow velocities ranging from a couple of hundred to a couple
of thousand  $\rm km \ s^{-1}$ relative to systemic. Moreover, in
some cases multiple velocity components have been identified
(\cite{coll01}; \cite{kaspi02}; \cite{kaastra02}). 
There are typically fewer features in emission than
absorption (e.g. \cite{kaastra02}; \cite{kaspi02}).
UV absorbing gas in Seyfert 1s was first observed with \iue (\cite{ulrich88}).
Multiple discrete kinematic components
to the UV absorbers have been  observed and it
has been suggested that the X-ray absorber is associated 
 with with one or more, but not necessarily all,  of the UV components 
(e.g. \cite{mathur95}; \cite{mathur99}).
Observations with the improved sensitivity and resolution
of \hst and \fuse have since shown that $\sim 60\%$ of Seyfert 1s exhibit
intrinsic UV absorption and that multiple kinematic
components are common (\cite{crenshaw99}).

At present, models of the X-ray
absorbers span a wide range in distance from the
central ionizing source, from
winds originating at the
accretion disk (\cite{elvis00}), out to the
putative (parsec-scale) molecular torus
(e.g. the multi-temperature wind model of Krolik \& Kriss (2001),
and beyond, to the NLR (e.g. Ogle \etal 2001).
In addition, for two particular Seyfert~1 galaxies observed
by \xmm (\mcg and Mrk~766) it has been proposed that
relativistically broadened soft X-ray lines from an accretion disk
can account for some of the spectral features
traditionally attributed to a warm absorber (Branduardi-Raymont \etal 2001;
Sako \etal 2001a).
However, Lee \etal (2001) have argued that the \chandra grating data, for
\mcg at least, can be modeled with a dusty warm absorber
without the relativistic emission lines. One thing is clear however:
all of the models must stand up to the scrutiny of an increasingly
large body of results as the results of new observational
campaigns become available. The purpose of the present paper
is to present the results of one such new campaign, namely
simultaneous \chandra \hetg, \hst, and \rxte observations of the luminous
($L_{2-10 \rm \ keV}$ typically $1.3-2.6 \times 10^{44} \ \rm ergs \ s^{-1}$
\footnote{We use $H_{0}=70 \ \rm km \ s^{-1} \ Mpc^{-1}$ and $q_{0}=0$
throughout this paper, unless otherwise stated.}, \cite{weav2001})
Seyfert~1 galaxy, \mk ($z=0.0344$, Fisher \etal 1995). 
\src has been studied
extensively in the UV and by every major X-ray astronomy mission
since {\it HEAO-1~A2}. Being so bright and exhibiting
interesting absorption structure in the UV
(e.g. \cite{kriss00}) and X-ray bands (e.g. Pounds \etal 1994;
\cite{reynolds97};
\cite{george98}; \cite{pounds01}; \cite{perola00}) \src
makes an excellent candidate for this kind of study.
The driving principles behind our
campaign were to measure,
with the highest spectral resolution available,
the X-ray and UV absorption features
{\it simultaneously} in order to eliminate uncertainty
due to variability, and to measure the hard X-ray continuum
{\it simultaneously} with the highest throughput
available (i.e. with \rxte) in order to compensate for the
poor efficiency of the \chandra gratings at energies above
$\sim 2$ keV. In the present paper we focus on the
soft X-ray spectroscopy results; detailed
results from the UV data are presented in a companion
paper (Kraemer \etal 2002, hereafter Paper~II). 
Our campaign was also designed to measure the
narrow and broad components of the Fe-K line and associated
Compton-reflection continuum, but these results are reported
elsewhere (\cite{yaqoob02b}).

The paper is organized as follows.
In \S\ref{sec:obs} we present the data and describe the analysis
techniques. In \S\ref{sec:overall} we discuss gross features
of the X-ray spectrum, including the intrinsic continuum form and
interpret the data in the context of historical, lower spectral
resolution CCD data. In
\S\ref{sec:features} we qualitatively
discuss the discrete X-ray spectral features before describing
detailed spectral modeling.
In \S\ref{sec:modelling} we describe in detail the modeling of
the X-ray spectrum using the photoionization code, XSTAR.
In \S\ref{sec:compare} we compare our results with
those from a previous \xmm grating observation
by Pounds \etal (2001). In \S\ref{sec:uv} we discuss the relationship
between the X-ray and UV absorbers.
Finally, in \S\ref{sec:conclusions} we summarize our
conclusions. 

\section{Observations and Data}
\label{sec:obs}
 
We observed \mk
with \chandra (simultaneously with \hst and \rxte)
on 2001 April 13--14
for a duration of $\sim 59$~ks,
beginning at UT 08~01:31.
The \chandra data were reprocessed
using {\tt ciao 2.1.3} and
{\tt CALDB} version 2.7, according to
recipes described in
{\tt ciao 2.1.3} threads\footnote{http://asc.harvard.edu/ciao2.1/documents\_threads.html}.
The \rxte PCA data were reduced using methods described in
Weaver, Krolik, \& Pier (1998), except that a later version of the spectral response matrix
(v 7.10) was used. A complete description of the \rxte observation and
data reduction is given in (\cite{yaqoob02b}) and will not be discussed further
here.
 
For \chandra
the instrument used in the focal plane of the
High Resolution Mirror Assembly (HRMA) was the High-Energy
Transmission Grating (or \hetg --
Markert, \etal 1995). The \hetg consists of two grating assemblies, a High-Energy
Grating (HEG) and a Medium-Energy Grating (MEG).
The approximate bandpasses of the HEG and MEG are $\sim 0.8-10$ keV and
$\sim 0.5-10$ keV respectively, although the usable portion
of these depends on the flux of the source.
Events dispersed by the gratings are collected
by a CCD array and can be assigned  an energy based on the
position along the dispersion axis. Since the CCDs have
intrinsic energy resolution, background events can be rejected
with a high efficiency, and different spectral orders can be easily
discriminated.
Genuine photon events collected by the CCDs fall into
specific pixel patterns classified by their {\it grade} and
we retained only grades 0, 2, 3, 4, and 6.
We utilized only the first-order grating data since the
zeroth-order data are piled up and the higher orders
carry much less counts than the first orders.
Therefore, only the summed, negative and positive first order \chandra grating spectra
were used in our analysis.
The mean \chandra total HEG and MEG count rates
were $0.2134 \pm 0.0015$
and $0.4813 \pm 0.0022$ cts/s respectively.
We accumulated separate HEG and MEG grating spectra from events along the
dispersion direction, and within  $\sim \pm 3.6$ arcseconds of the
peak in the cross-dispersion direction.
The source flux showed little variability over
the entire duration of the campaign. For example,
for HEG plus MEG lightcurves binned at 1024 s,
the excess variance above the
expectation for Poisson noise (e.g. see Turner \etal 1999)  was $(-0.8 \pm 3.7) \times 10^{-4}$,
consistent with zero.
HEG, MEG and PCA spectra were therefore
extracted over the entire on-time for each instrument.
This resulted in net exposure times of 57,950 s
for
HEG and MEG (which includes a deadtime factor of 0.0129081), and
80,624 s for the PCA.
 
We made effective area files (ARF, or {\it ancillary response file})
using {\sc ciao}~2.1.3,
which also take account of the dithering of the
satellite. Photon spectra
(summed over the $\pm 1$ orders)
were made by correcting the counts spectra
for the effective area and cosmological
redshift, but not Galactic absorption.
We also extracted counts spectra in a form suitable
for use by the spectral-fitting package XSPEC (version 11.0.1)
\footnote{http://heasarc.gsfc.nasa.gov/docs/xanadu/xspec/}.
Again, we used first-order events only, combining the 
positive and negative orders
but keeping the HEG and MEG spectra separate for independent analysis.
The spectra
were produced with a range of different bin sizes,
from $0.02 \AA$ to $0.64 \AA$. The smaller bin sizes
are appropriate for studying narrow spectral features
which may be unresolved, and the larger bin sizes
are better for studying broad features and the continuum.
For reference,
FWHM spectral resolutions of the HEG and MEG are
$0.012 \AA$ and $0.023 \AA$ respectively.
Since the MEG
soft X-ray response is much better than the HEG
(whose bandpass
only extends down to $\sim 0.8$ keV) we will
use the MEG as the primary instrument but refer to the HEG
for confirmation of features in the overlapping bandpass and
for constraining the continuum.
The MEG spectral
resolution corresponds to FWHM velocities of $\sim 280, 560$, and
3,560~$\rm km \ s^{-1}$ at observed energies of 0.5, 1.0, and 6.4 keV respectively.
The HEG spectral resolution corresponds to velocities of about half of
these values.
The response matrices
{\tt acismeg1D1999-07-22rmfN0004.fits} and
{\tt acisheg1D1999-07-22rmfN0004.fits}, combined with the ARF files
described above, were used to fold models through the instrument response
and thereby directly compare predicted and observed counts spectra.

We did not subtract detector or X-ray background since it is
such a small fraction of the observed counts
($<0.5\%$ for the MEG).
We treated the statistical errors on both the photon and counts
spectra with particular care since the lowest and highest energies
of interest can be in the Poisson regime, with spectral bins often
containing a few, or even zero counts.
We assign a statistical error of $1.0 + \sqrt{(N+0.75)}$ on the
number of photons, $N$, in a given spectral bin. This
prescription, from Gehrels (1986),
is a good approximation to the positive error for a
Poissonian distribution, being good to better then $1.5\%$ for all $N$,
but over-estimates the negative error for
small $N$. As $N$ increases, the approximation tends to the Gaussian
limit. It is important to bear in mind however, that even for
100 counts per bin, $\sqrt{N}$ still gives a statistical error
which is $\sim 10\%$ too small (the corresponding error on the
approximation at this $N$ is $<0.1\%$ -- see Gehrels 1986).
We were also careful not to propagate statistical errors when
combining spectral orders or binning spectra. Statistical
errors were always computed on numbers of photons in  final bins,
according to the above prescription, since the standard method of
propagation of errors is only correct when the errors are Gaussian.

For the HEG data, the signal-to-noise ratio per $0.02\AA$ bin
is $> 1$ blueward of  $\sim 14\AA$ and 
as high as $\sim 7$ in the $\sim 2-3.5\AA$ range. 
For the MEG data, in the $\sim 2-25\AA$ band which we examine
here,
the signal-to-noise ratio per $0.02\AA$ bin ranged from $\sim 1$
at $\sim 20\AA$ to a maximum of $\sim 12$ at $\sim 7\AA$.
The systematic uncertainty in the energy scale is currently
believed to be $0.0028\AA$ and $0.0055\AA$ for the HEG and
MEG respectively\footnote{http://space.mit.edu/CXC/calib/hetgcal.html}.
For the MEG, at 0.5 keV, 1 keV and 6.4 keV this corresponds
to velocity offsets of $\sim 67$, 133, and 852~$\rm km \ s^{-1}$ respectively, and
about half of these values for the HEG.

\section{Overall Spectrum and Preliminary Spectral Fitting}
\label{sec:overall}

We used XSPEC v11.0.1 for spectral fitting to the HEG and MEG spectra
in the 0.8--5 keV and 0.5--5 keV bands respectively.
These energy bands will be used in all the spectral fitting in the
present paper, in which we concentrate on 
features in the soft X-ray spectrum (less than $\sim 2$ keV). 
We used the $C$-statistic for finding the best-fitting model
parameters, and quote 90\% confidence, one-parameter statistical errors 
unless otherwise stated.
The harder spectrum, out to 19 keV, (using simultaneous
\rxte data in addition to the \chandra data) has been discussed
at length in Yaqoob \etal (2002b). The hard X-ray continuum 
measured by \chandra MEG, HEG, and \rxte was found to be
well described with a single power law in the range 2--19 keV,
with no Compton-reflection continuum required,
and a photon index of $\Gamma = 1.674^{+0.008}_{-0.009}$,
measured from the joint three-instrument spectral fits.
In all the spectral fits in the present paper we shall include
this hard power-law component for the continuum, with $\Gamma$
fixed at 1.674. 
Detailed analysis of the Fe-K emission
complex was also presented by Yaqoob \etal (2002b) so we shall not
discuss this part of the spectrum further. Since we convolve all
models through the instrument response before comparing with the
data, all our model line-widths are intrinsic values and do
not need to be corrected for the instrument broadening. This
contrasts with the analysis method in some of the literature
on \hetg results in which models are fitted directly to the
data and {\it then} corrected for instrument broadening 
(e.g. Kaspi \etal 2001, 2002).

First, we show how the 0.5--5 keV MEG data compare to a
simple model consisting only of the hard ($\Gamma = 1.674$) power law
and Galactic absorption. For the latter, we use a value of
$4.44 \times 10^{20} {\rm cm^{-2}}$ (\cite{murphy96}) throughout this
work.
\figufspecnhgal (a) shows the MEG photon spectrum of \mk 
binned at $0.32\AA$ with the model overlaid. 
The bottom panel in
\figufspecnhgal (b) shows the ratio of the data to this model. 
This simple power law is clearly a poor fit to the data, with considerable
complexity apparent, both in the continuum, and in terms of
discrete absorption lines and possible edges.
\figufspecnhgal clearly shows a soft
X-ray excess, rising up from the hard power law
(modified by Galactic absorption), below $\sim 0.7 \ {\rm keV}$,
reaching $\sim 60\%$ above the extrapolated hard power-law
spectrum. 
We caution that there may still be broad residual uncertainties of
$\sim 30\%$ or more in the calibration of the MEG effective area
at the lowest energies
\footnote{http://space.mit.edu/CXC/calib/hetgcal.html}. 
To quantify this further, we have 
examined \chandra \hetg data for 3C~120
(observed in 2001, December;
details will be reported elsewhere).
The
residuals when the MEG data are fitted with a power-law plus
Galactic absorption are less than 20\% in the entire 0.5--5 keV
band. This strongly suggests that at least part of the soft excess in \mk is
real. 
This soft excess has been observed during simultaneous
\rosat/ \ginga observations (Pounds \etal 1994), a
\bsax observation (Perola \etal 2000), and a 
recent \xmm observation (Pounds \etal 2001). Both the 
\bsax and \xmm 
observations required a photoionized absorber as well as 
a soft excess. A 1994 \asca observation on the other
hand, did not require a soft excess in addition to the
complex absorber (Reynolds \etal 1997; George \etal 1998).

The origin of the soft excess is unclear. It may be the tail end of 
some kind of soft thermal emission (or Comptonized soft
thermal emission), possibly from an accretion disk 
(e.g. see Piro, Matt, \& Ricci (1997), 
and references therein). Any relativistically-broadened
\oxla emission that is present could also contribute to the
soft excess (e.g. Branduardi-Raymont \etal 2001; Turner \etal 2001).
Since the soft excess appears only in the 0.5--1 keV band
of our data, we do not have enough information to constrain
its origin, and sophisticated modeling of it is not warranted.
Therefore, in the remainder of this paper we model the
0.5--5 keV intrinsic continuum with a broken power law
in which the hard photon index, $\Gamma_{1}$, above a break energy, $E_{B}$,
is always fixed at 1.674 and the soft photon index, $\Gamma_{2}$ (below $E_{B}$),
is a free parameter, as is $E_{B}$.
When the data are modeled with this continuum, there is
still considerable structure in the soft X-ray spectrum,
indicative of the presence of a photoionized, or `warm' absorber.
Now, the warm absorber in \mk and 
other AGN when observed with CCDs (such as those on \asca)
has often been modeled simply with absorption edges
due to \oxyseven and \oxyeight since the lower CCD spectral
resolution did not usually warrant more sophisticated models.

In order to directly compare the new \chandra data with
previous observations we modeled the MEG data with
two absorption edges and the intrinsic continuum described 
above. The energies and optical depths (at threshold) of
the two absorption edges were allowed to float.
The best-fitting model overlaid on the MEG photon spectrum
is shown in \figufspecedges. 
The best-fitting parameters obtained from this
model were
$0.732^{+0.004}_{-0.019} \ {\rm keV}$ and $0.871^{+0.014}_{-0.020} \ {\rm keV}$
for the threshold energies of \oxyseven and \oxyeight
respectively. For the the optical depths at threshold
we obtained  $\tau = 0.30^{+0.04}_{-0.04}$ and
$0.14^{+0.05}_{-0.03}$ for 
\oxyseven and \oxyeight respectively.
All quantities refer to the rest frame of \mk.
The best-fitting soft X-ray photon index was 2.06 and the
break energy was 1.28 keV.
The measured edge energies are in good agreement
with the expected values (0.739 keV and 0.871 keV for
\oxyseven and \oxyeight respectively), although the
\oxyseven edge is marginally inconsistent at a 
confidence level of 90\% or greater. We can compare
the threshold optical depths with those measured
from the 1994 \asca observation (Reynolds 1997):  
$0.11^{+0.04}_{-0.03}$ (\oxyseven) and $0.04^{+0.04}_{-0.04}$
(\oxyeight). The MEG values are larger
(although they are
consistent with lower
signal-to-noise measurements 
from \bsax data (Perola \etal 2000),
which has much lower spectral resolution).
However, it is known
that the apparent optical depths of the \oxyseven and \oxyeight 
edges are variable in
some AGN (e.g. Otani \etal 1996; Guainazzi \etal 1996).
Also, since the spectrum is so complex around
the regions of the \oxyseven and \oxyeight edges,
optical depths derived from simple models should
be interpreted with caution. We will show in \S\ref{sec:modelling}
that photoionization models can adequately describe the
data with smaller values of the \oxyseven and \oxyeight edge depths.
Even the heavily binned spectrum in Fig.~2
shows a discrete feature at 0.922 keV (rest frame)
corresponding to Ne~{\sc ix} resonance absorption. This causes
the \oxyeight edge to be somewhat deeper than that which would
be obtained if the  Ne~{\sc ix} resonance absorption were not
present. Additionally, Fe~M-shell 
unresolved transition arrays (UTAs), which have been
observed in \src and some other sources
(e.g. Pounds \etal 2001; Sako \etal 2001b; Lee \etal 2001; Kaspi \etal 2002;
Kaastra \etal 2002), may at some level affect the
inferred column densities of both \oxyseven and O~{\sc viii}.

These factors which affect the apparent \oxyseven and
\oxyeight edge depths obtained from simple modeling
clearly have an impact on the puzzling
results from variability studies. For example, 
Otani \etal (1996) found that in the Seyfert~1  galaxy
MCG~$-$6$-$30$-$15, the \oxyeight edge depth varied
(in response to the continuum) whilst the \oxyseven
edge depth remained constant. In contrast, Guainazzi \etal
(1996) found that the \oxyseven edge was variable whilst
the \oxyeight edge depth remained constant.
For MCG~$-$6$-$30$-$15, Morales, Fabian, \& Reynolds (2000) proposed
a solution consisting of  multiple, spatially
separated zones to explain these results. The higher
spectral resolution \chandra data shows that these
issues need to be revisited.
Indeed, as has been pointed out by
Lee \etal (2001), more recent \chandra observations of
\mcg suggest that the \oxyeight edge in
this source is complicated by \nenine resonance absorption, as 
well
as a complex of lines from \feseventeen and \feeighteen.

\section{Spectral Features and Kinematics}
\label{sec:features}

\subsection{Absorption and Emission Lines}
\label{sec:abslines}
In order to best illustrate different characteristics of
the spectrum we display the spectral data in several
different ways. In \figufspeccont we show the MEG photon
spectrum below $\sim 1.4$ keV, as a function of observed energy, along with
the deduced intrinsic broken power-law continuum
(not corrected for Galactic absorption), as derived from the
the best-fitting photoionization model described in \S\ref{sec:modelling}.
The solid red curve in \figufspeccont is the  
photoionization model itself (the details of
which are deferred to \S\ref{sec:modelling}).
The data and the models in \figufspeccont are all absorbed
by the Galactic column density.
An important point to realize from \figufspeccont is that
in some regions the spectrum is so complex that it is
difficult to distinguish absorption from emission. For
example, at observed energies between $\sim 0.7-0.8$ keV,
there appears to be a complex emission feature relative to the
local two-edge model continuum in \figufspecedges but from \figufspeccont
it is clear that the peak flux of this feature
(between $\sim 0.78$--0.79 keV, observed) coincides with
the inferred intrinsic continuum, implying that the apparent
feature is simply due to the presence of broad absorption
complexes on either side of it. 
Also, there is a sharp variation in the MEG effective area
at $\sim 0.8$ keV and any uncertainties here could manifest 
themselves 
as apparent spectral features (see \S\ref{sec:modelling} for
further discussion).

In \figmegsnr and \fighegsnr we show the MEG and HEG spectra
respectively, this time as a function of wavelength
in the source rest-frame, redward of $2\AA$.
These spectra have a bin size of $0.02\AA$, approximately
the FWHM MEG spectral resolution ($0.023\AA$).
Below each panel showing the data is the signal-to-noise
ratio (SNR) in each bin, so that one can easily guage the statistical
significance of any spectral features. The same MEG data
are also displayed in \figmeglyman and \figmegbalmer.
In \figmeglyman we overlay the Lyman series (blue)
for H-like N, O, Ne, Mg, Si, S, Ar and the corresponding He-like 
triplets (red) that lie in the 2-25$\AA$ range.
In \figmegbalmer we overlay the He-like
resonance series lines (blue) to the $n=1$ level for the same elements,
and in addition the Balmer lines (red) of the
hydrogenic ions of Si, S, and Ar.
 
Referring to Figs. 3 to 7, we 
see that 
the strongest, clear discrete spectral features detected are
absorption lines due to
\nila, \nilb, \oxla, \neniner \resonetwo ($\lambda 13.447\AA$), \neniner 
\resonethree ($\lambda 11.549\AA$),
\nela, and \mgelevenr \resonetwo ($\lambda 9.169\AA$). Some strong Fe absorption 
lines are also detected. 
Weaker absorption lines are also apparent from the spectra, such
as higher order Lyman lines of O and Ne
(there is evidence for transitions up to $n=5$ in each case).
Higher signal-to-noise X-ray spectroscopy of NGC~3783 (Kaspi
\etal 2002) has shown that blending of absorption features
considerably complicates the spectrum. In particular,
lines from \feseventeen, \fenineteen, and \fetwentyone
are likely to be blended with the absorption
troughs
at 
$\sim 11.5,12.1,13.4,16.0\AA$ in the \mk spectrum.
We note that we do not detect absorption from 
the highest ionization states of elements beyond Ne, like
\sila (see \figmeglyman). Thus, the ionization state of the
absorber cannot be so high that Si~{\sc xiv} is abundant.

Apart from \fekalfa and some tentative emission features
at $\sim 9.3,9.6,9.8, 12.6\AA$ (possibly attributable to
\fenineteen), no other emission-line
features are clearly identifiable. 
In particular, we do not detect the \oxyseven He-like triplet,
which {\it was} detected during an \xmm observation, by 
Pounds \etal (2001). In \S\ref{sec:compare} we shall give a more
detailed comparison with the \xmm observation.

The presence of broad, bound-free absorption features
and many narrow features which may be blended,
makes it impossible to define an observed continuum
for much of the spectrum,
and therefore 
difficult to measure meaningful equivalent widths of
even the strongest narrow absorption features.
Rather than try to measure equivalent widths 
of {\it all} candidate features, and compare
with model predictions, we will take the physically more
direct approach of fitting photoionization models to the data
(\S\ref{sec:modelling}).
Nevertheless, in the process of measuring the observed
energies of the strongest discrete absorption features,
we did attempt to measure the equivalent widths of {\it some} absorption lines.
The measurements were made using the
broken power law plus two-edge model as the baseline
spectral model (see \S\ref{sec:overall}) and Gaussians
to model the absorption lines. The continuum and edge parameters
were frozen at the best-fitting values given in
\S\ref{sec:overall}, and the data were fitted over
a few hundred eV centered on
the feature of interest. The fitted range of the data was
extended down to 0.47 keV in order to measure the
parameters of \nila. 
Initially the {\emph{intrinsic}} width of each Gaussian
was frozen at less than the instrumental resolution but then
allowed to float in order to investigate whether the lines
were resolved. The results for the
measured energies, apparent velocity shifts relative
to systemic, equivalent widths, and 
intrinsic widths, of these strongest absorption features 
are given in Table~1. Note that the quoted equivalent
widths are relative to the {\it intrinsic} continuum. 
Again, we caution that the measured equivalents widths are subject
to uncertainties in the continuum, line-blending, and any complexity
in the line profile. It is instructive to compare
line strengths with a physical model so, also given in
Table~1 are the predicted equivalent widths from the
best-fitting photoionization model discussed in \S\ref{sec:modelling}. 

\subsection{Absorption-Line Velocity Profiles}
\label{sec:profiles}

Table~1 shows that all of the measured absorption lines
are blueshifted, with Gaussian-fitted centroids ranging
from near systemic velocity to $\sim -600$~$\rm km \ s^{-1}$. However,
some of the velocity profiles appear to be more complex
than Gaussian, as can be seen from \figvelprof
(which shows the profiles of
\nela, \neniner \resonetwo ($\lambda 13.447\AA$),   
\mgelevenr \resonetwo ($\lambda 9.169\AA$), and \oxla).
They are centered with zero at the systemic velocity
of \src, with positive and negative velocities
corresponding to redshifts and blueshifts respectively. 
The velocity 
profiles
were obtained from combined HEG and MEG data except for the \oxla profile
which was
determined using MEG data only, since at this energy the HEG effective
area is
negligible. 
It can be seen from Table~1 and \figvelprof that the
velocity offset for all of the absorption lines
is consistent with $-200$~$\rm km \ s^{-1}$, except for the
the \mgelevenr resonance line. The latter has an apparent
minimum at $\sim -600$~$\rm km \ s^{-1}$ which 
may indiciate
a truly different absorption-line profile, or the different
offset velocity may be an artifact of contamination
from other absorption lines, or 
there may be emission filling in a trough
at $-200$~$\rm km \ s^{-1}$.

Table~1 also shows that \nela and \nenine \resonetwo 
($\lambda 13.447\AA$) lines appear to be marginally resolved,
whilst all the other lines are unresolved
(FWHM$<300$~$\rm km \ s^{-1}$ for the lightest elements,
going up to FWHM$<1030$~$\rm km \ s^{-1}$ for Mg).
The \nela line  is particularly interesting because 
we obtain a different centroid offset velocity
depending on whether the Gaussian model has an
intrinsic width fixed at a value much less
than the instrument resolution, or whether the
intrinsic width is a free parameter. The latter
gives a marginally better fit ($\Delta C = -4.0$),
with an offset of $\sim -210$~$\rm km \ s^{-1}$, and
a broad line, but still with FWHM~$<1330$~$\rm km \ s^{-1}$.
The narrow-line fit gives a higher blueshift, but
in both cases the lower limit on the velocity offset
is the same, at $\sim -400$~$\rm km \ s^{-1}$. This is indicative
of a complex velocity profile.
However, the MEG FWHM velocity resolution
goes from $\sim 280$~$\rm km \ s^{-1}$ for the lowest-energy
transitions (Nitrogen lines),
up to $\sim 730$~$\rm km \ s^{-1}$ for the
highest-energy transition (\mgelevenr) so any
apparent structure in the profiles should be interpreted
with caution. The resolution at \nela is 
550~$\rm km \ s^{-1}$, FWHM.
The complexity in the velocity profiles, if real, may either be
intrinsic or due to contamination from other
absorption and/or emission features.
The intrinsic complexity could be due, for example,
to the presence of several kinematic
components, making different contributions to the overall
profile, as is the case for UV lines (see \S\ref{sec:uv}).

\figvelprof~(a) shows that the
broad trough in the \nela profile, centered at 
$\sim -200$~$\rm km \ s^{-1}$, extends
redward of the systemic velocity,
perhaps by up to $\sim +400$~$\rm km \ s^{-1}$.
\figvelprof~(b) is centered around the
\neniner \resonetwo ($\lambda 13.447\AA$) transition and the velocity profile
appears to be much more asymmetric than \nela, having
a clear, single minimum at $\sim -200$~$\rm km \ s^{-1}$. The profile of
\neniner ($\lambda 13.447\AA$) appears to be just as broad as \nela, however.
\figvelprof~(c) shows the velocity profile for \mgelevenr \resonetwo 
($\lambda 9.169\AA$)
which appears to have two minima, one at $\sim -600$~$\rm km \ s^{-1}$ and the
other at systemic. It is possible that the profile is really
the same as that of \neniner ($\lambda 13.447\AA$), with the trough
contaminated by emission. The velocity profile of \oxla
shown in \figvelprof~(d), is very similar to that of \mgelevenr 
($\lambda 9.169\AA$),
with two minima at the same velocities, except that the
contrast of the minima with the center of the profile is not
so high.

\figvelproftwo shows velocity spectra
centered on several atomic transitions that are {\it not} 
significantly detected in
the data (\nala, \arla, \mgla, \sila). The
abundances of Na and Ar are small compared to Oxygen and Neon
for example, partially accounting the absence of features
due to these ions. 
There is a hint of \mgla absorption but it is
very weak if present.
\sila is clearly not present in absorption in the
data, although there is a hint of emission. 
Since the cosmic abundances of Silicon and Magnesium are similar,
and
since we see \mgelevenr but we do not see \sila, this 
constrains the ionization state of the absorber and
will therefore be important for 
photoionization modeling, discussed in \S\ref{sec:modelling}.

The velocity profiles in \figvelprofone are 
reminiscent of 
\chandra observations of absorption features in other AGN.
The
absorption is generally blueshifted with respect to the systemic velocity and
there is evidence of multiple kinematic components
and multiple ionization states 
(e.g. NGC~4051, Collinge \etal 2001; NGC~3783, Kaspi \etal 2002; 
NGC~5548, Kaastra \etal 2002, Yaqoob \etal 2002a). 
Due to the limited spectral resolution of \chandra, it is 
possible that there are actually many unresolved kinematic
components making up the velocity profiles and we shall
return to this point in \S\ref{sec:uv} where we examine
the relationship between the X-ray and UV absorbers.

\section{Photoionization Modeling}
\label{sec:modelling}

We used the photoionization code XSTAR 2.1.d to generate several grids of
models of emission and absorption from photoionized gas in
order to directly compare with the data. In Paper~II CLOUDY was
used to model the UV data. We used the default
solar abundances in XSTAR
and CLOUDY.
For reference, these abundances are given in Table~2.

\subsection{The Spectral Energy Distribution}
\label{sec:sed}
First, we constructed a spectral energy distribution (SED) for \mk
as follows.
Average radio and infrared fluxes were obtained
from Ward \etal (1987).
UV fluxes were obtained from our \hst spectrum,
observed simultaneously
with {\it Chandra}, from 
continuum regions centered on 1181, 1366, 1510, 2291, and
$3013 \AA$ (observed frame). The fluxes were dereddened using the Galactic
reddening curve of Savage \& Mathis (1979) and a reddening value of E(B$-$V)
$=$ 0.08, corresponding to the observed H~I Galactic column
(Shull \& Van  Steenberg 1985).
Further details of
the UV data analysis can be found in Paper~II. 
In principle
the X-ray portion of the SED must be derived iteratively because
the intrinsic X-ray continuum can only be deduced from fitting the
data self-consistently with a model of the absorption (which is
complex and affects the X-ray data over a broad energy range,
up to nearly 2 keV). In practice we implemented a two-step
process to obtain a fiducial SED and then investigated
the effects of adjusting this SED in a manner described below.
For the first of the two steps, we started with an SED in which the
X-ray spectrum was a single power law with a photon index of
1.674 (consistent with \chandra / \rxte data above 2 keV; see
Yaqoob \etal 2002b), extrapolated down to 0.5 keV. The 0.5 keV point
was then simply joined onto the last UV point by a straight line
in log-log space (this corresponds to a
slope of $-1.44$ in $L_{\nu}$ versus $\nu$). 
The hard X-ray power law in this and subsequent
SEDs extended out to 500 keV. Grids of photoionization models
(details of which are given below) were then generated
using this initial SED and the \chandra MEG and HEG data were
then fitted below 5 keV using a broken power law model for the intrinsic
continuum, modified by the photoionized absorber and Galactic absorption.
The hard power-law index was kept fixed at $\Gamma_{1} = 1.674$.
We obtained a soft power-law index of $\Gamma_{2} = 2.19$ and a break
energy of 1.05 keV.
In the second step, we used the broken power-law, with the above
parameters
(and normalization obtained from the spectral fit), for the X-ray
portion of the SED, down to 0.5 keV. Again, we connected
the 0.5 keV point to the last UV point by a straight line
in log-log space. The resulting SED is shown in \figsed
(filled circles connected by solid lines). The dashed curves show
variations
on this SED, constructed by raising and lowering the flux at 55 eV
(the energy required to doubly ionize He)
by a factor 1.5. These latter two SEDs will be used to
investigate the effects of uncertainties in the 
unobserved part of the SED by
comparing with the results obtained with the baseline SED.
Also shown in \figsed is the `mean AGN' SED derived
by Matthews \& Ferland (1987).

\subsection{Model Grids}
\label{sec:model}
The photoionization model grids used here are two-dimensional,
corresponding to a range in values of
equivalent neutral Hydrogen column density, $N_{\rm H}$,
and the ionization
parameter, $\xi = L_{\rm ion}/(n_{e} r^{2})$ erg  cm $s^{-1}$.
Here  $L_{\rm ion}$ is the ionizing luminosity (in $\rm erg \ s^{-1}$),
in the range 1 to 1000
Rydbergs,
$n_{e}$ is the electron density in ${\rm cm^{-3}}$,
and $r$ is the distance of the
illuminated gas from the ionizing source in cm
\footnote{In Paper~II a different ionization parameter is used,
$U \equiv Q_{\rm ion}/(4\pi n_{e} c r^{2})$. Here $Q_{\rm ion}$
is the number of ionizing photons above 1 Rydberg,
$c$ is the speed of light, and $n_{e}$ and $r$ have the same
meaning as for $\xi$. From our baseline SED, the conversion
factor is $U = 0.03208 \xi$, or $\log{U} = \log{\xi} -1.49$.
Therefore, our best-fitting value of
$\log{\xi} = 1.76$ corresponds to $U=1.85$, or $\log{U}=0.27$.}.
The grids were computed for equi-spaced intervals
in the logarithms of $N_{\rm H}$ and  $\xi$, which ranged
from $5 \times 10^{20} - 5 \times 10^{22} \ {\rm cm^{-2}}$
and $0.5-2.5$ respectively. The X-ray data cannot
constrain the electron density since we do not
detect any suitable emission lines, so we fixed
$n_{e}$ at $10^{8} \ {\rm cm^{-3}}$ in most of the grids.
For diagnostic purposes, 
some grids were constructed with $n_{e}$ as low as
$10^{2} \ {\rm cm^{-3}}$ and as high as $10^{11} \ {\rm cm^{-3}}$.
We confirmed that results from fitting the
photoionizied absorber models to the X-ray
data were indistinguishable for densities in the
range 
$n_{e}=10^{2} \ {\rm cm^{-3}}$ to $10^{11} \ {\rm cm^{-3}}$.
Hereafter all XSTAR models discussed correspond to
a density of $10^{8} \ {\rm cm^{-3}}$, unless otherwise stated.
This value is arbitrary and does not signify any preference
for high densities.

Finally, all models
assumed a velocity turbulence ($b$-value) of $1$~$\rm km \ s^{-1}$, and 
since XSTAR broadens lines by the greater of the
turbulent velocity and the thermal velocity, it is
the latter which is relevant for these models.
As the turbulent velocity increases, the equivalent
width of an absorption line increases for a given column
density of an ion. 
Therefore, to calculate model equivalent widths
correctly one must know what the velocity width is
(note that the $b$-value could represent all sources of
line-broadening). We cannot simply extend the XSTAR model
grids into another dimension (velocity width), and deduce a
$b$-value directly from model-fitting, because we are limited by the
finite internal energy resolution of XSTAR.
The widths of the energy bins
of the XSTAR model grids are in the range
$\sim 200-600$~$\rm km \ s^{-1}$.
Our choice of velocity width ($b$-value corresponding
to the thermal width) ensures that the
energy resolution of XSTAR is the limiting factor and not
the $b$-value itself. We will then need to take a rather
more complicated approach to model-fitting, which we describe
in \S\ref{sec:modelfits} below.

\subsection{Model-Fitting}
\label{sec:modelfits}

We proceeded to fit the MEG and HEG spectra (binned at
$0.08\AA$)
simultaneously, with all
corresponding model parameters for the two instruments 
tied together, except for
the relative normalizations. The intrinsic continuum was
a broken power law (the hard photon index again fixed at 1.674)
modified by absorption from photoionized gas and the Galaxy.
Excluding overall normalizations, there were a total
of four free parameters, namely the column density of the
warm absorber, $N_{H}$, the ionization parameter, $\xi$,
the intrinsic soft photon index, $\Gamma_{2}$, and the
break energy of the broken power-law continuum, $E_{\rm B}$.

Our model-fitting strategy is more complicated than 
simply comparing the XSTAR model spectra with the data since
the XSTAR spectra assume a certain line width ($b$-value), which
itself is limited by
the finite internal energy resolution of XSTAR.
We chose $b= 1 \ \rm  km \ s^{-1}$ in order that the energy resolution of XSTAR
is the limiting factor. The result is that the equivalent
widths of lines in the XSTAR spectra need to be 
calculated more rigorously
using a line width ($b$-value) appropriate for the data.
On the
other hand, the ionic columns in the XSTAR models are robust from
this point-of-view and do not depend on the velocity width.
Therefore, after finding the best-fitting ionization parameter
and warm-aborber column density
from a global fit using the model described above, in \S~\ref{sec:comparison} 
we will deduce a $b$-value from a curve-of-growth analysis, using
ionic column densities from the best-fitting global model. We will
then  
construct detailed model profiles for each of the detected
absorption lines.

The best-fitting parameters derived from the XSTAR global model
fits were
$N_{H} = 2.06^{+0.39}_{-0.45}\times 10^{21} \ {\rm cm^{-2}}$ and
$\log{\xi} = 1.76^{+0.13}_{-0.14}$ ergs cm ${\rm s^{-1}}$ (or $\log{U}=0.27$). The broken
power
law model corresponding to this best fit had a break energy of
$E_{\rm B}=0.95^{+0.15}_{-0.09} \ {\rm keV}$ and a soft power-law 
photon index of
$\Gamma_{1} = 2.36^{+0.21}_{-0.20}$, values which are consistent with the
SED that was used as input to the models. 
\figxstaroverdata shows the best-fitting model
folded through the MEG response and overlaid
onto the MEG counts spectrum.
The lower panel of \figxstaroverdata
shows the ratio of the
data to the best-fitting model. \figxstarmodel shows the actual
best-fitting XSTAR model before folding through the 
instrument response.
\figxstarcontour shows the 68\%, 90\%, and 99\% joint
confidence contours of $\log{\xi}$ versus $N_{H}$.
Overall, the fit
is very good and it is already apparent that
the best-fitting parameters which describe the
overall spectrum also give good fits to the
most prominent absorption lines and edges.
This can also be seen in \figufspeccont which shows
the best-fitting model overlaid on the photon spectrum,
and in \figvelprofone and \figvelproftwo which show close-ups of the data
and model (in velocity space),
centered on particular atomic transitions.
We note that formally, a second solution with a higher column
density ($\sim 3 \times 10^{21} \ {\rm cm^{-2}}$) but
similar ionization parameter is also viable, but the
best-fitting break energy of the intrinsic continuum is
at $\sim 2.15 \pm 0.25$ keV, inconsistent with the input SED
and with the measured value for the first solution.
We can therefore eliminate this second solution.

The main driver of these broadband fits is the overall
shape of the observed spectrum, but the
narrow absorption lines, and the
absorption edges also affect the fits.
Since the centroids of the absorption lines are blueshifted,
a velocity offset had to be applied to the XSTAR model.
Although the velocity offsets may be different for
different absorption lines (see \S\ref{sec:profiles}),
the signal-to-noise of the data is such that anything
more complex than a single velocity offset for all the
absorption lines is not warranted.
We therefore applied a nominal velocity offset of $-200$~$\rm km \ s^{-1}$
relative to systemic,
which was satisfactory for all of the absorption features.
We had to apply an additional (but non-physical)
correction of $-270$~$\rm km \ s^{-1}$. The reason for the latter offset
is that the code XSTAR v2.1.d contains hard-wired wavelengths of 
some atomic transitions which disagree 
with wavelengths
published elsewhere 
\footnote{For example, in the {\it Atomic Line List}
(at http://www.pa.uky.edu/~peter/atomic/).}. 
We found that a single, simple offset of $-270$~$\rm km \ s^{-1}$ was
required and \figufspeccont shows that this gives an excellent
fit for all the detected absorption lines (reported in Table~1),
except possibly \mgelevenr \resonetwo ($\lambda 9.169\AA$). The latter has negligible
impact on the spectral fit, given the signal-to-noise of the
data. The fact that the
individual model line profiles give good
fits to the data (\figvelprofone), and were calculated using the
ionic column densities from the best-fitting XSTAR model, 
published wavelengths, and the deduced $b$-value (see \S\ref{sec:comparison}), shows
that the various assumptions are good for our purpose.

We mentioned in \S\ref{sec:overall} that the 2.0--2.5 keV region
in the \chandra spectra suffers from systematics as large as
$\sim 20\%$ in the
effective area due to limitations in the calibration of
the X-ray telescope absorption edges in that region
\footnote{http://asc.harvard.edu/udocs/docs/POG/MPOG/node13.html}.
We therefore
omitted the  2.0--2.5 keV regions of the \chandra spectra
during the spectral fitting so that the 
fits would not be unduly biased.
These and other instrumental absorption edges can be seen
in the upper panel of \figxstaroverdata.
All the MEG data in the
0.5--5 keV band are shown here, although the 2.0--2.5 keV
region was omitted during spectral fitting. Other regions
which have sharp changes in the effective area also
show notable mismatches between data and model, but 
these are not as bad as in the 2.0--2.5 keV region. In particular,
the region between $\sim 0.76-0.81$ keV shows an
apparent broad emission feature.
However, the entire region in the $\sim 0.7-0.9$ keV is
spectrally very complex (see \figufspeccont). 
We also mentioned in \S\ref{sec:abslines} 
that the peak of the apparent emission between
0.78--0.79 keV (observed) actually
coincides with the intrinsic continuum (\figufspeccont)
so could be the result of a lack of absorption rather than
an emission feature. However we see that our best-fitting
XSTAR model actually predicts {\it absorption} due to
\oxyeight Ly$\gamma$ and \feseventeen (at an observed energy
of $\sim 0.78-0.79$ keV, or $\sim 0.81-0.82$ keV rest-frame).
This implies that there must be some emission features in
the data. We were not able to model this emission with
the emitted spectra from the XSTAR models,
even with $\xi$ and $N_{H}$ unconstrained 
(without over-predicting other features which are not observed).
Note that there is also emission (above the intrinsic 
continuum) at 0.80 keV (0.828 keV rest-frame); 
\feseventeen and \feeighteen lines are possible candidates.
However, we should remember that the region of the spectrum
around $\sim 0.8$ keV does coincide with a
sharp change in the MEG effective area (\figxstaroverdata) 
so it is likely that the poor fit 
here may in part be due to
calibration uncertainties. 
Nevertheless, we confirmed
that omitting the data in the region
$\sim 0.76-0.81$ keV had a negligible effect on the
derived best-fitting model parameters, so we decided to retain
the data in this region since the uncertainties
are certainly not as large as those in the 2--2.5 keV region.

\subsection{Detailed Comparison of Data and Photoionization Model}
\label{sec:comparison}

In this section we compare in detail the best-fitting XSTAR 
photoionized absorber model and the data (refer to 
\figufspeccont for the spectrum, and to \figvelprofone and \figvelproftwo
for some `velocity spectra'). Note that although the best-fitting
model was derived using spectra binned at $0.08 \AA$, we can compare
the model and data at higher spectral resolution, depending on the
signal-to-noise of the feature in question.
As discussed in \S\ref{sec:abslines} the strongest absorption-line
features in the \chandra data are \nila, \nilb, \oxla, \neniner \resonetwo 
($\lambda 13.447\AA$), \nela,
\neniner \resonethree ($\lambda 11.547\AA$), and \mgelevenr \resonetwo 
($\lambda 9.169\AA$) (see also Table~1).
No
strong, identifiable
emission features were found below 5 keV. \figufspeccont shows
that the best-fitting XSTAR model described above
($N_{H} = 2.06 \times 10^{21} \ \rm cm^{-2}$, $\log{\xi}=1.76$
[or $\log{U}=0.27$])
predicts the strength of all these absorption lines
very well, despite the fact that in this
{\it direct} comparison the model equivalent widths
have not been computed for the correct line width
($b$-value), as explained at the end of \S\ref{sec:modelfits}. 

In order to determine what the appropriate $b$-value might
be, we calculated curves-of-growth 
for different values
of $b$ (see \figcog). We can see from Table~1 that
all the Ly~$\alpha$ transitions from H-like
and He-like ions have approximately the same measured
equivalent widths, within errors.
This suggests that all of these lines are on the saturated part of the 
curve-of-growth.
The theoretical curves-of-growth in
\figcog indicate that the measured equivalent widths of $\sim 10^{-3} \AA$
are all consistent with
$b \sim 100 \ \rm km \ s^{-1}$. This is a rather model-independent
conclusion since this deduction does not utilize any information
from the XSTAR modeling. We also plotted the measured equivalent
widths from Table~1 (converted to Angstroms) as
a function of ionic column density (from the best-fitting XSTAR
model described above) and overlaid these on the
curves-of-growth (\figcog). This again illustrates that,
for solar abundances,
all the measurements are consistent with $b=100$~$\rm km \ s^{-1}$ 
(except possibly \oxla). The predicted equivalent widths
for $b=100$~$\rm km \ s^{-1}$ are shown in Table~1
so that they can be compared directly with the observed 
equivalent widths. 

The velocity profiles in \figvelprofone (for four of the
strongest absorption lines: \nela, \neniner ($\lambda 13.447\AA$),
\mgelevenr ($\lambda 9.169\AA$), \& \oxla) also
show very good agreement between data and model.
In this case the model curves were calculated using the
broken power-law and two-edge continuum described in \S\ref{sec:overall},
a Gaussian for the absorption line, with 
$\sigma = b/\sqrt{2} = 70.7$~$\rm km \ s^{-1}$,
and an equivalent width from Table~1.
\figvelproftwo illustrates that some features not
predicted by the model are indeed not detected in the data
(as would be expected from the low ionic columns in Table~3).
These include \arla, \nala, \mgla and \sila. These are all higher
ionization
features (and Na and Ar have relatively low abundance) so we can see that the
absence of features from these ions in the data contributes to the strong constraints on the
ionization parameter.
Thus, our single-component photoionization model gives
surprisingly good fits to the principal absorption lines.

There are many more discrete absorption features apparent in the
spectra but with lower signal-to-noise.
Although these individual weak features cannot be studied in 
detail we can at least check whether the best-fitting XSTAR model 
does not conflict with the data. We found that, as far as
narrow absorption lines are concerned, all absorption lines
predicted by the model were consistent with the data, with the
exception of the 0.78--0.79 keV (observed) region, which
has already been discussed in \S\ref{sec:overall} and
\S\ref{sec:abslines} (see \figufspeccont). 
In particular, we found that the model agrees 
well with the data for the low-order Lyman transitions of Ne and
O. The first five \oxyseven $1s-np$ resonance lines also 
agree well with the data (the rest are too weak). 
Conversely, the data show possible absorption features which
are not predicted by the model, but none of these are statistically
significant.

\figufspeccont also shows that the
best-fitting XSTAR model is
good at the \oxyseven edge but
could be better if the edge were deeper since the flux
just above the edge is somewhat over-predicted.
We note that the Fe ionic 
columns predicted by the best-fitting XSTAR model are largest for 
ionization states \fefourteen to \fenineteen
(see Table~3). According to 
Behar, Sako, \& Khan (2001), this should yield an
Fe~M-shell unresolved transition array
(UTA) in the energy region of interest.
This has in fact has been detected by \xmm in \mk (Pounds \etal 2001),
albeit at lower energy
(centered at $\sim 0.77$ keV, rest-frame; 0.744 keV, observed-frame),
corresponding to lower ionization states dominated by Fe~{\sc xi-xii}.
Such UTAs have also been detected in some other Seyfert~1 galaxies
(IRAS~13349$+$2438, Sako \etal 2001b; MCG~$-$6$-$30$-$15,  Lee \etal 2001;
NGC~3783, Kaspi \etal 2002; NGC~5548, Kaastra \etal 2002).
Fe~L-shell absorption could also be present
near the O~{\sc vii} (and the  O~{\sc viii} edge)
due to neutral Fe locked
up in dust (see \figufspeccont) and
this is not modeled by XSTAR. Such a scenario was proposed by Lee \etal
(2001) for \mcg. However, we do not see evidence
for these Fe~L edges in \mk and  this is consistent with the
amount of reddening, which corresponds to 
a threshold optical depth of the neutral Fe~L3 edge of $\sim 0.02$,
not detectable by the MEG. 
The 
complex of \oxyseven $1s^{2}-1snp$ resonance transitions with
$n>5$ may also complicate the region just above
the O~{\sc vii} edge (see also Lee \etal 2001).
We also point out that the data around $\sim 1.02-1.04 \ {\rm keV}$ 
are not well-fitted by the model
(see \figufspeccont). However the break energy of the broken power
law is at 1.04 keV (observed frame), and the break in the model spectrum is rather
sharp so a more physical model of the continuum may
fit better. 

The best XSTAR model absorber fit 
around the O~{\sc viii} edge and the 
\neniner ($\lambda 13.447\AA$) absorption line (which is not far blueward
of the O~{\sc viii} edge) is much better than the
fit in the O~{\sc vii} edge region. In \S\ref{sec:overall} we
mentioned how spectral fits 
using simple absorption-edge models (representing O~{\sc vii} and O~{\sc viii}) 
could be misleading.
We see that
more sophisticated modeling with better data
now gives a smaller O~{\sc viii} edge since the
data above the O~{\sc viii} edge are affected
by other absorption features (\neniner ($\lambda 13.447\AA$) in particular) 
which can be accounted for by the photoionization models. 
This has also been pointed out by Lee \etal (2001).
However, the edge depth may still be over-estimated due
to the presence of Fe inner shell UTAs.
From the XSTAR models we can extract the
\oxyseven and \oxyeight ionic column densities
at the extremes of the $90\%$ confidence intervals
for $N_{H}$ versus $\xi$.
We find that the
\oxyseven column is $3.5^{+2.7}_{-1.7}\times 10^{17} \ {\rm cm^{-2}}$
and the \oxyeight column is
$8.1^{+1.1}_{-2.1}\times 10^{17} \ {\rm cm^{-2}}$.
Given an absorption cross-section at threshold of
$2.75(1.09) \times 10^{-19} \ \rm{cm^{-2}}$ for \oxyseven(\oxyeight)
(\cite{verner95}), we find that the model optical depths
(at threshold) of the O edges are
$\tau_{{\rm \oxyseven}}=0.10^{+0.07}_{-0.05}$ and
$\tau_{{\rm \oxyeight}}=0.09^{+0.01}_{-0.02}$ respectively. These
values
can be compared with those obtained from a \mk \asca
observation in 1994 in which simple `edge-models' yielded
$\tau_{{\rm \oxyseven}}=0.11^{+0.03}_{-0.04}$ and
$\tau_{{\rm \oxyeight}}=0.04^{+0.04}_{-0.03}$
(\cite{reynolds97}; see also George \etal 1998).
Whilst the \oxyseven optical depth is consistent, the
\oxyeight optical depth is only marginally consistent.
However, variability of the warm absorber has
been observed in several AGN and \mk may be no exception.

We investigated the robustness of the XSTAR model fits to the absorption
features by examining the data versus model
at the extremes of the $N_{H}$ versus $\xi$
confidence intervals (see \figxstarcontour). 
By picking pairs of values of $N_{H}$ and $\xi$ in the
plane of the contours, we can freeze the column density and
ionization parameter at the chosen values and find the
new best-fit for each pair of values (this is in fact
how such contours are constructed). Then we can examine
the data and model spectra in detail for each of these
new fits to see where the model and data deviate 
significantly from each other.
We found that 
the models were still able to account for the
strengths of all of the strong absorption-line features
(listed in Table~1) for pairs of
$N_{H}$ and $\xi$ lying inside the 68\% confidence contour.
On the other hand, for pairs of  $N_{H}$ and $\xi$ 
lying near the 99\% contour we found
that the \nela absorption line was particularly
sensitive and the equivalent width would be over-predicted
or under-predicted by as much as a factor of two.
Therefore, we conclude that the confidence contours in \figxstarcontour
are conservative, and the true parameter space
for the allowed range of $N_{H}$ and $\xi$ is even more
restricted than the 99\% confidence contours in \figxstarcontour
indicate.

\subsection{Emission-Line Spectra }
\label{sec:emlines}

As well as absorption,
the XSTAR code also calculates emission spectra assuming
a shell of gas subtending $4\pi$ steradians at the central ionizing
source. The fact that we do not detect any strong, identifiable emission
features 
does not place constraints directly on the covering
factor since we do not know the gas density or its
distance from the ionizing source. Requiring that the
shell of gas is much thinner than its distance to the source
(i.e. $\Delta r/r \ll 1$), and using the best-fitting
values of column density and ionization parameter,
only constrains the upper limit on the absorber 
distance on the order of kpc (e.g. see Turner \etal 1993).

It has been argued in the literature, based on \xmm RGS 
grating data, that the soft X-ray spectra of \mcg and 
Mrk~766 can be better explained by relativistically-broadened
emission lines due to \csix~Ly$\alpha$, \nila, and \oxla rather than
a pure photoionized absorber (e.g. Branduardi-Raymont \etal 2001; however,
see Lee \etal 2001 for counter-arguments).
Due to the different MEG bandpass, we can only test for
relativistically-broadened \oxla emission in our \mk data.
Firstly, we note that we cannot obtain even
a reasonable spectral-fit with only a relativistic disk
line and no warm absorber. When we added an emission-line
model from a relativistic
disk around a Schwarzschild black hole (e.g. Fabian \etal 1989)
to the best-fitting photoionized absorber model, we obtained
a reduction in the $C$-statistic corresponding to
only 90\% confidence for the addition of two free
parameters (the disk inclination and line intensity).
The inner and outer radii of the disk were fixed
at 6 and 1000 gravitational radii respectively, and the
radial line emissivity per unit area was a power law
with index $-2.5$. The line energy was fixed at
0.65362 keV in the \mk rest frame. The best-fitting
inclination angle and equivalent width was $41^{+4}_{-8}$ degrees and
$17 \pm 9$ eV respectively. Pounds \etal (2001) 
assumed an
inclination angle of $30^{\circ}$ when fitting ionized disk 
reflection models, and in Paper~II we also argue for a 
small inclination angle.
If we fix the disk
inclination at $30^{\circ}$ in the \chandra fit, 
we find a much smaller equivalent width of 
$8^{+8}_{-6}$~eV. 
When the \oxla emission line is included in the model,
the best-fitting XSTAR model of the absorber does
not change much: $\log{\xi}$ decreases from
1.76 to 1.65 and $N_{H}$ decreases by $\sim 13\%$ to
$1.80 \times 10^{21} \ \rm cm^{-2}$.
We caution that in our case
the broad line may simply be modeling residuals which
are remaining calibration uncertainties and/or curvature
in the spectrum, and conclude that the evidence
for relativistically-broadened \oxla emission in our data is not
compelling but not ruled out either. 

\subsection{Effects of Changes in the SED}
\label{sec:deltaSED}
To investigate the effect of uncertainty in the energy range
$\sim 10-500 \ {\rm eV}$ in our fiducial SED, we generated XSTAR models
using
alternative SEDs in the manner described for the fiducial SED
(see \S\ref{sec:sed}). Two
alternative SEDs were used, in which the flux
at
$55 \ {\rm eV}$ was either 50\% lower or 50\% higher relative
to the fiducial SED. These SEDs
are illustrated by dashed lines in \figsed. 
Proceeding with model-fitting just as described for the
fiducial SED, we found that the best-fitting column density
($N_{H}$), ionization parameter ($\xi$), soft power-law photon index,
and broken power-law break energy,  were 
consistent for all three SEDs, within the statistical errors.
Moreover, the models with the alternative SEDs were able to
provide fits to all the major discrete absorption features
and edges that were just as good, or no worse, than the
fiducial SED models. We also found that  
the predicted ionic columns of \oxyseven and \oxyeight for the best-fitting 
models using the alternative SEDs did not differ by more than 10\% from
the values obtained with the baseline SED model.
Thus, our results and conclusions are not sensitive to any 
reasonable uncertainties in the unobserved EUV part of the
SED. 

\section{Comparison with an \xmm Observation in October 2000}
\label{sec:compare}

\mk was observed by \xmm on 2000 October 25 (\cite{pounds01}), $\sim 180$
days before our \chandra \hetg observation. In this section
we compare our \chandra results to the \xmm results for
the soft X-ray spectrum of \mk. The Fe-K region and
higher energy spectrum has already been discussed in detail, and
compared to the \xmm results, in Yaqoob \etal (2002b).
Several factors need to be borne in mind when making the
comparison of the soft X-ray spectra. 
First, the effective areas of the \xmm gratings
(RGS) exceed that of the \chandra MEG below $\sim 1.15$ keV.
Second, the spectral resolution of the RGS at 0.5 keV is
at best $\sim 850$~$\rm km \ s^{-1}$ (compared to $\sim 280$~$\rm km \ s^{-1}$ for the
MEG). The spectral resolution in both cases is worse at
higher energies. Third, the exposure times for the \xmm CCDs
were $\sim 25-27$ ks, and $\sim 30$ ks for the RGS. This is
to be compared to $\sim 58$ ks for the \chandra exposure time.
Fourth,
the 0.5--10 keV observed flux during the
\xmm observation was $2.6 \times 10^{-11} {\rm \ ergs \ cm^{-2} 
\ s^{-1}}$, a factor of 2.8 less than the corresponding MEG 0.5--10 keV flux of 
$7.3 \times 10^{-11} \rm \ ergs \ cm^{-2} \ s^{-1}$.

\xmm and \chandra observed a similar underlying continuum. 
The EPIC \xmm CCD data yielded a 2--10 keV power-law index
of $1.66 \pm 0.02$, in good agreement with our \chandra/ \rxte
measurements of $1.674^{+0.010}_{-0.015}$ (\cite{yaqoob02b}). 
A strong soft excess relative to this power law was
observed by both \xmm and \chandra and with a similar
magnitude relative to the hard power law. 
Pounds \etal (2001) report detecting only two significant
absorption features, namely a blend of \neniner and \fenineteen
at $13.52 \AA$, and a broad absorption trough, 
likely to be associated with an  Fe M-shell 
unresolved transition array (UTA), centered 
around $16.1\AA$. According to the calculations of
Behar \etal (2001), this corresponds to ionization states
dominated by Fe~{\sc xi-xii}. 
The \neniner ($\lambda 13.447\AA$)
feature was detected by the MEG, and our data are not inconsistent with 
the Fe~M-shell UTA.
We established the latter by adding a Gaussian
absorption trough to our best-fitting XSTAR model, in order
to model the UTA (using the parameters measured by
Pounds \etal 2001), and found that the agreement between data and
model 
improved and the data did
not deviate by more than one standard deviation from the model
in the 0.70--0.78 keV region (observed). 
At the \nenine/ \fenineteen feature 
the underlying continuum was a factor of $\sim 2$ higher 
during the \chandra observation than in the \xmm observation.
The equivalent widths  of the \nenine/\fenineteen absorption
features ($1.0^{+0.5}_{-0.6}$ eV for \chandra, $2.2 \pm 1.0$ eV
for \xmm) are compatible with the absorption
responding to the change in continuum level, as would
be expected. 
Since Pounds \etal (2001) do not give any further measurements or
limits (in particular on the \oxyseven and \oxyeight edges),
or results of photoionization modeling, it
is difficult to make a more detailed comparison of the
warm absorber during the two observations.  

Pounds \etal (2001) also detected \oxyseven He-like 
triplet emission in the RGS data, with the intercombination
line being the strongest (equivalent width $1.5 \pm 0.7$ eV).
Smaller equivalent widths of the \oxyseven He-like
triplet emission lines in the \chandra MEG data  
would be expected if
the intensity of the emission lines remained constant
whilst the underlying continuum increased (by more than
a factor of two in this case). None of the \oxyseven triplet
lines are detected significantly, most likely due to
insufficient signal-to-noise. For example, at the
energy of the \oxyseven \resonetwo resonance line,
the combined effective area of the two RGS instruments on
\xmm is a factor $\sim 60$ higher than the MEG. Combined
with the shorter exposure and lower source flux for
the \xmm observation, the signal-to-noise of the
RGS data at the \oxyseven \resonetwo resonance line
is a factor $\sim 3$ higher than in the MEG data. 
Nevertheless, we do
obtain a non-zero intensity from Gaussian modeling of the
MEG data.
Direct measurement of the intercombination
line in the MEG data gives an equivalent width of
$1.00^{+0.44}_{-0.65}$ eV. Unfortunately, the
signal-to-noise is insufficient to determine whether
the emission-line intensity did or did not respond to the
change in the continuum, so we cannot constrain the
distance of the emitter.

\section{Relation Between the  X-ray and UV Warm Absorbers}
\label{sec:uv}
 
Our \chandra \hetg observation was simultaneous with an
\hst observation. Full details of the results from the UV
data are discussed in Paper~II.
Here we simply summarize very briefly the findings of Paper~II
and discuss whether the UV absorption predicted
by our simple model of the X-ray absorber is consistent with the UV data.
 
The \hst spectrum of \mk shows multiple kinematic
components to the UV absorption features, a characteristic which
is found to be common in Seyfert~1 galaxies (e.g. Crenshaw \& Kraemer 1999).
Specifically, eight distinct kinematic components are
identified for
\cfour, \nfive, and \sifour.
The seven principal kinematic components (numbered 1 through 7 in Paper~II) lie
at $-422, -328, -259, -62, -22, +34, +124 \ \rm km \ s^{-1}$ respectively, 
relative to systemic. 
\figuvvelocities shows the UV absorber velocities 
superimposed on some of the strongest absorption features in the \chandra \hetg data,
along with the best-fitting XSTAR 
photoionization model profiles. Shown are the
velocity profiles of \nela, \neniner \resonetwo ($\lambda 13.447\AA$), 
\mgelevenr \resonetwo ($\lambda 11.547\AA$), and
\oxla.
The UV kinematic components are clumped into two groups
and together clearly
span the widths of the X-ray profiles.
It is possible these
X-ray velocity profiles actually consist of several
kinematic components which cannot be resolved by \chandra.
We convolved the UV kinematic components through the MEG
response function and found the resulting profiles
to be
compatible with the observed X-ray profiles.
The observed X-ray profiles of \neniner \resonetwo ($\lambda 13.447\AA$) and 
\mgelevenr \resonetwo ($\lambda 9.169\AA$) appear to be
somewhat different to those
of \nela and \oxla, the former pair being somewhat asymmetric. 
As pointed out in \S\ref{sec:features}
and \S\ref{sec:modelling}, this could be due to blending with unresolved
line emission. Alternatively, the profiles could be intrinsically different.
However, we again caution that the velocity resolution of the
X-ray profiles shown in \figvelprof is between $\sim 350$--730~$\rm km \ s^{-1}$ FWHM so the differences
in the X-ray profiles must be regarded as tentative. 

Although the X-ray and UV absorbers may share the same 
velocity space,
in Paper~II we show that the ionization state and column
densities of the UV absorbers are too low to produce the
observed X-ray absorption. Photoionization models
for each of the UV kinematic components yielded
column densities, $N_{H}$,
in the range $1.77 \times 10^{18} \rm \ cm^{-2}$ to
$4.00 \times 10^{19} \rm \ cm^{-2}$ and
$\log{\xi}$ in the range 0.01 to 1.33
(or $\log{U}$ in the range $-1.48$ to $-0.16$).
Conversely, the column density and ionization parameter
of the X-ray absorber is high.
The best-fitting parameters from
photoionization modeling of the X-ray data are
$\log{\xi}=1.76$ ($\log{U}=0.27$) and 
$N_{H} = 2.06 \times 10^{21} \ \rm cm^{-2}$.
Some UV absorption is
predicted by the best-fitting XSTAR model to the \chandra data
but the relevant ionic column densities are small enough that the
predictions of the X-ray model are not in conflict with the UV data.
Table~3 gives some column densities of some key ions from our
best-fitting photoionization model to the X-ray data, compared
with some measured column densities from the UV data, as well
as some corresponding predicted column densities from the UV photoionization
modeling. Indeed, it can seen that 
in the X-ray absorber, the predicted column densities of
C~{\sc ii}, C~{\sc iii},  C~{\sc iv}, N~{\sc v}, and O~{\sc vi}
could easily be hidden in the UV absorbers.
Full details, including a plausible physical and geometrical
picture of the X-ray and UV absorbers can be found in Paper~II.

\section{Conclusions}
\label{sec:conclusions}
 
We observed \mk 
for $\sim 59 \ {\rm ks}$ with the \chandra \hetg on 2001 April 13--14,
simultaneously with
\rxte and \hst.
The complex Fe-K line 
and Compton-reflection
continuum are discussed in
Yaqoob \etal (2002b).
Details of the UV data
and results from the \hst observation are given in Paper~II.
Here we summarize the main results from the \chandra
\hetg soft X-ray spectrum, in view of
relevant constraints from the \rxte and
\hst observations.
 
\begin{enumerate}
\item{Combined \chandra MEG, HEG and \rxte data show that the hard X-ray
spectrum of \mk is well described by a simple power law, with little Compton
reflection continuum required, and a photon index of
$\Gamma = 1.67$ above 2 keV.
At lower
energies, below $\sim 1$ keV, the
spectrum steepens, the soft excess
attaining a magnitude of about 60\% higher than
the extrapolated hard power law.
When the intrinsic X-ray
continuum is modeled with a broken power-law, self-consistently
modeling the X-ray absorption with a photoionization model
(see below)
yields
a break energy
$E_{B}=0.95^{+0.15}_{-0.09} \ {\rm keV}$ and a soft power law index of
$\Gamma=2.36^{+0.21}_{-0.20}$.
During the observing
campaign the broadband source flux was $5.1 \times 10^{-11} \rm \ ergs \
cm^{-2} \ s^{-1}$, typical of
historical values (the corresponding 2--10 keV luminosity was
$1.3 \times 10^{44} \rm \ ergs \ s^{-1}$).}

\item{Below $\sim 2$ keV, the {\it observed} spectrum
shows considerable deviations from a smooth continuum,
due mainly to bound-free absorption opacity, and complexes
of discrete absorption features.
Particularly noteworthy 
are the \oxyseven and \oxyeight edges at
$0.732^{+0.004}_{-0.019} \ {\rm keV}$ and $0.871^{+0.014}_{-0.020} \ {\rm keV}$
respectively. Within the errors,
both values are consistent
(\oxyseven marginally so) with their respective rest energies, and 
from the best-fitting photoionization models the
optical depths at threshold are $0.10^{+0.07}_{-0.05}$ for \oxyseven
and $0.09^{+0.01}_{-0.02}$ for O~{\sc viii}. Compared
with a previous \asca observation (\cite{reynolds97}; \cite{george98}), 
the \oxyseven edge is
consistent, whilst the \oxyeight edge is somewhat larger
in the MEG data. 
However, we note that 
the edge depths 
may be variable in some AGN
(e.g. Otani \etal 1996; Guainazzi \etal 1996).
Estimates of the \oxyeight edge depth are
also complicated by nearby absorption features due,
for example, to Ne and Fe.
The spectral region between the
\oxyseven and \oxyeight edges is very complex and the most difficult
to model with a simple photoionized absorber. This part of the spectrum
is also where one would expect Fe~L edges (e.g. due to dust) and
Fe~M-shell unresolved transition arrays (UTA), neither of which are detected.
The data, are, however consistent with a UTA with model parameters
measured during an \xmm observation (Pounds \etal 2001).}
 
\item{On the next level of detail,
we detect absorption lines from H-like ions of N, O, Ne, He-like
ions of O, Ne, Mg, as well as absorption features due to highly ionized Fe. The
ionization state of the absorber is high, but not so high that absorption
due to H-like Mg or Si is detected. Only the
\nela and \nenine \resonetwo lines
appear to be marginally resolved, and the rest are unresolved
(FWHM$<300$~$\rm km \ s^{-1}$ for N, going up to FWHM$<1330$~$\rm km \ s^{-1}$ for Mg).
The absorption lines are consistent with an
outflow of $-200$~$\rm km \ s^{-1}$ relative to systemic.
However, the detailed velocity profiles of the absorption features 
are not all the same for all ionic species.
The complexity and differences in profiles could be due
to the contribution from different, unresolved
kinematic components and/or 
blending and contamination from absorption/emission due to other
atomic transitions.}

\item{We have modeled the \chandra \hetg spectra using the 
photoionization code XSTAR 2.1.d,
with a spectral energy
distribution (SED) constructed using our UV and X-ray data.
The best-fitting ionization parameter and neutral
equivalent Hydrogen column density of the absorbing gas are
$\log \xi=1.76^{+0.13}_{-0.14}$ ergs cm ${\rm s^{-1}}$ (or $\log{U}=0.27$), and
$N_{H}=2.06^{+0.39}_{-0.45}\times 10^{21} {\rm cm^{-2}}$ respectively.
This best-fitting model gives an excellent fit to
the overall {\it observed} spectrum. It also is able to
model all the principal local absorption features to a degree
ranging from fair to excellent, except for the region around
0.78--0.80 keV. The poor fit here may be, in part,
due to uncertainties in the instrument response function.
A curve-of-growth analysis indicates that a
velocity width of $b=100 \ \rm km \ s^{-1}$ gives a good
match between the measured absorption-line equivalent widths.
Our results are insensitive to reasonable
uncertainties in the unobserved EUV part of the SED.}
 
\item{We do not detect any
clearly identifiable emission lines (apart from the Fe-K line),
in contrast with an earlier \xmm observation in
which Pounds \etal (2001) detected He-like \oxyseven
triplet
emission, which constrained the
electron density to be $>10^{11} \rm \ cm^{-3}$. However,
this lower limit is subject to uncertainties in the
line-ratio measurements. 
Although the broadband X-ray flux during that observation was
nearly a factor of three lower than it was during our campaign,
the signal-to-noise of our
data is worse, and insufficient to constrain the distance of
the emitter from the ionizing source. Since we cannot assume that
the emitter and absorber have the same ionization state and/or
column density, time-resolved spectroscopy with better signal-to-noise
is required to address the
question of the location of the absorber and emitter.
Deducing the mass outflow of the X-ray absorber must also await
better data because the density and global covering factor are
unknown.}

\item{Simultaneous \hst observations of \mk indicate that there are
eight kinetic components comprising the UV absorbers, with
velocities ranging from $-422$ to $+124$~$\rm km \ s^{-1}$.
The X-ray absorber is compatible with sharing the same 
velocity space
as the UV absorbers.
It is possible that the velocity profiles of the X-ray absorbers
are also made up of several kinematic
components, possibly the same as the UV ones,
but unresolved because of the factor of $>30$ worse
velocity resolution of the X-ray  data compared to the UV data.
The X-ray profiles do however hint at complex structure,
possibly two kinematic components, or groups of components.
The UV components certainly appear to cluster into two groups.
There are also differences in the X-ray velocity profiles
between some of the ionic species. This may either be due
to genuine differences in gas dynamics but may also
be due to unresolved and unmodeled line emission. On the other hand,
other Seyfert~1 galaxies also show two kinematic components
(or groups of components) in their absorption-line profiles,
but with different offsets relative to systemic
(NGC~5548, \cite{kaastra02}; NGC~4051, \cite{coll01};
NGC~3783, \cite{kaspi02}).}
 
\item{Although the X-ray and UV absorbers may share the same 
velocity space,
Paper~II shows that the ionization state and column
densities of the UV absorbers are too low to produce the
observed X-ray absorption. Conversely, the ionization and
column density of the X-ray absorber is high; some UV absorption is
predicted by our best-fitting photoionization model to the \chandra data
but the relevant ionic column densities 
can be hidden in the UV absorbers.}
 
\end{enumerate}

The authors gratefully acknowledge support from
NASA grants NCC-5447 (T.Y.), NAG5-10769 (T.Y.), NAG5-7385
(T.J.T), NAG5-4103 (S.B.K.), and CXO grant GO1-2101X (T.Y., B.M.).
This research
made use of the HEASARC online data archive services, supported
by NASA/GSFC and also  
of the NASA/IPAC Extragalactic Database (NED) which is operated by 
the Jet Propulsion Laboratory, California Institute of Technology, 
under contract with NASA.
The authors are grateful to the \hst, \chandra and \rxte
instrument and operations teams for making these observations
possible, and to Tim Kallman for much advice on XSTAR.
The authors also thank Julian Krolik
for useful discussions, and an anonymous referee for doing a very
thorough job on the manuscript. TY would like to dedicate this
paper to his mother, Zubaida Begum Yaqoob, who passed away in March 2001,
after sacrificing so much in order that her children could get an
education and pursue their goals.

\newpage
\begin{deluxetable}{lrrrr}
\tablecaption{Absorption Lines in the Chandra HETGS Spectrum of \mk}
\tablecolumns{5}
\tablewidth{0pt}
\tablehead{
\colhead{Line $^{a}$} & \colhead{EW (Data)}$^{b}$ & \colhead{EW (Model) $^{c}$}
& \colhead{Velocity $^{d}$ } & \colhead{FWHM} \nl
& \colhead{(eV)} & \colhead{(eV)} & \colhead{($\rm km \ s^{-1}$)} & \colhead{($\rm km \ s^{-1}$)} \nl}
\startdata

\nila & $0.55^{+0 e}_{-0.15}$ & 0.33 & $-290^{+130}_{-90}$ & $<310$ \nl

\nilb & $0.64^{+0e}_{-0.26}$ & 0.67 & $-230^{+100}_{-100}$ & $<320$ \nl

\oxla & $0.72^{+0e}_{-0.13}$ & 0.94 & $-85^{+200}_{-135}$ & $<450$ \nl

\neniner \resonetwo ($\lambda 13.447\AA$) & $1.33^{+0e}_{-0.30}$ & 1.08 & $-220^{+260}_{-165}$ & $440^{+590}_{-305}$ \nl

\neniner  \resonethree ($\lambda 11.547\AA$) & $1.12^{+0.20}_{-0.68}$ 
& 0.87 & $-180^{+280}_{-250}$ & $<890$ \nl 

\mgelevenr \resonetwo ($\lambda 9.169\AA$) & $0.68^{+0.48}_{-0.50}$ 
& 1.05 & $-620^{+330}_{-180}$ & $<1040$ \nl 

\nela $^{f}$ & $1.89^{+0.74}_{-1.27}$ & 0.96 & $-210^{+155}_{-205}$ &
$875^{+460}_{-550}$ \nl

\nela $^{f}$ & $1.08^{+0.29}_{-0.39}$ & 0.96 & $-355^{+205}_{-60}$ 
& 1 (fixed) \nl
\enddata
\tablecomments{\small Absorption-line parameters measured
from the MEG spectrum, using simple Gaussians
(see \S\ref{sec:features}).
All measured quantities refer to intrinsic parameters,
already corrected for the instrument response.
Errors are 90\% confidence for one interesting parameter
($\Delta C = 2.706$). All velocities have been rounded
to the nearest 5~$\rm km \ s^{-1}$.
$^{a}$ Laboratory-frame wavelengths. $^{b}$ Measured
equivalent widths in the \mk frame. $^{c}$ Predicted equivalent widths
using $b=100 \ \rm km \ s^{-1}$ and 
XSTAR columns (\S\ref{sec:comparison}).
$^{d}$ Velocity offset (\mk frame) of Gaussian
centroid relative to systemic. Negative values are blueshifts.
$^{e}$ No meaningful upper limits due to poor statistics and line
saturation.
$^{f}$ Different centroids were obtained for \nela depending
on  whether the intrinsic width of the Gaussian was
fixed at 1~$\rm km \ s^{-1}$ or a free parameter, indicating a complex
profile (see \S\ref{sec:features} and \figvelprofone).} 
\end{deluxetable}

\newpage

\begin{deluxetable}{lr}
\tablecaption{Element Abundances used in XSTAR and CLOUDY models}
\tablecolumns{2}
\tablewidth{0pt}
\tablehead{
\colhead{Element} & \colhead{Abundance} 
}
\startdata
          H  & $1.00 \times 10^{0}$ \nl
          He & $1.00 \times 10^{-1}$ \nl
          C  & $3.54 \times 10^{-4}$ \nl
          N  & $9.33 \times 10^{-5}$ \nl
          O  & $7.41 \times 10^{-4}$ \nl
          Ne & $1.20 \times 10^{-4}$ \nl
          Mg & $3.80 \times 10^{-5}$ \nl
          Si & $3.55 \times 10^{-5}$ \nl
          S  & $2.14 \times 10^{-5}$ \nl
          Ar & $3.31 \times 10^{-6}$ \nl
          Ca & $2.29 \times 10^{-6}$ \nl
          Fe & $3.16 \times 10^{-5}$ \nl
          Ni & $1.78 \times 10^{-6}$ \nl
\enddata
\end{deluxetable} 

\newpage

\begin{deluxetable}{lrrrrr}
\tablecaption{ Ionic Column Densities ($10^{14} {\rm \ cm^{-2}}$)}
\tablecolumns{6}
\tablewidth{0pt}
\tablehead{
\colhead{Ion } & \colhead{X-ray  $^{a}$} &
\colhead{UV  $^{b}$} & \colhead{UV  $^{b}$ } & \colhead{Ion } & \colhead{X-ray  $^{a}$} \nl
& \colhead{(Predicted)} & \colhead{(Predicted)} & \colhead{(Measured)} &  & \colhead{(Predicted)}
}
\startdata
 
H~{\sc i} & $1.4 \times 10^{1}$ & 
$1.6 \times 10^{1}$ $^{c}$ & $4.7 \times 10^{1}$ & Mg~{\sc xi} & $3.0 \times 10^{2}$ \nl
C~{\sc ii} &$3.0 \times 10^{-7}$ & - & $<3.5 \times 10^{-1}$ & 
Mg~{\sc xii} & $4.9 \times 10^{1}$ \nl
C~{\sc iii} & $6.7 \times 10^{-4}$ &
$5.5 \times 10^{-1}$ & $<5.1 \times 10^{-1}$ & Si~{\sc xiv} & $3.8 \times 10^{0}$ \nl
C~{\sc iv} & $1.2 \times 10^{-1}$ & 7.3 & 7.3 & 
Ar~{\sc xviii} & $1.6 \times 10^{-3}$  \nl
N~{\sc v} & $8.5 \times 10^{-1}$ & 7.1 & 7.0 & \fetwelve & $8.4 \times 10^{0}$ \nl
O~{\sc vi} & $7.1 \times 10^{1}$ & $1.2 \times 10^{2}$ & 
$0.82 \times 10^{2}$ & \fefourteen & $1.0 \times 10^{2}$ \nl 
O~{\sc vii} & $3.5 \times 10^{3}$ & - & - 
& \fefifteen & $1.8 \times 10^{2}$ \nl
O~{\sc viii} & $8.1 \times 10^{3}$ & - & - & \feseventeen & $1.6 \times 10^{2}$ \nl
Ne~{\sc ix} & $1.4 \times 10^{3}$ & - & - & \feeighteen & $5.1 \times 10^{1}$ \nl
Ne~{\sc x} & $7.5 \times 10^{2}$ & - & - & \fenineteen & $7.6 \times 10^{0}$ \nl
\tablecomments{\small $^{a}$ X-ray column densities (in units of
$10^{14} \rm \ cm^{-2}$)
refer to those predicted by the best-fitting XSTAR
model to the \chandra data 
(see \S\ref{sec:modelling}).
The predicted column
densities for Si~{\sc iii} and Si~{\sc iv} from the XSTAR models were
negligible.
The upper limits on Si~{\sc iii} and Si~{\sc iv}
from summing over UV kinematic
components were $0.21$ and $0.07 \times 10^{14} \ {\rm \ cm^{-2}}$ respectively.
$^{b}$ Predicted and measured UV column densities are
summed over available measurements from all UV kinematic components except
the low ionization component of 4 (see Paper~II).
H~{\sc i}, C~{\sc iii} and O~{\sc vi} measurements are from
Kriss \etal (2000) and {\it not}
the present observations. Predicted UV column densities are
from CLOUDY photoionization models of the UV data.
$^{c}$ Column from all kinetic components except 4 (low) and
$4^{\prime}$ (see Paper~II).}
\enddata
\end{deluxetable}

\newpage

\section*{Figure Captions}
 
\par\noindent
{\bf Figure 1} \\
Comparison of the \mk MEG spectrum
to the
best-fitting power-law model fitted
to the $2-19 \ {\rm keV}$ (joint \hetg/ \rxte data)
extrapolated down to $0.5 \ {\rm keV}$.
Galactic absorption of $4.44\times 10^{20} {\rm cm^{-2}}$ is included in the
model and the power-law index is
$\Gamma=1.674$. 
The MEG data are binned at $0.32\AA$. The top panel shows the
photon spectrum and the bottom panel shows the
ratio of the MEG data to the above model. Both panels show that
the intrinsic spectrum is {\it required} to steepen at low energies,
relative to the hard X-ray power-law.
 
\par\noindent
{\bf Figure 2} \\
Comparison of the \mk MEG data (binned at $0.32\AA$) to 
a broken power-law model modified by two
absorption edges (representing opacity due
to \oxyseven and \oxyeight), and Galactic
absorption ($4.44\times 10^{20} {\rm cm^{-2}}$).
The hard X-ray photon index is fixed at $1.674$, and
the best-fitting soft X-ray index and break energy
are $2.06$ and $1.28 \ {\rm keV}$ respectively.
The best-fitting edge energies agree well with the expected 
rest-frame energies
of the \oxyseven ($0.739 \ {\rm keV}$) and \oxyeight ($0.871 \ {\rm keV}$) 
edges.
This kind of model has historically been used to model lower
spectral resolution CCD data.
Note that using this model, the inferred
optical depth of the \oxyeight edge could be
affected by the \nenine \resonetwo ($\lambda 13.447\AA$) absorption feature 
(at $\sim 0.89 \ {\rm keV}$, observed).
The estimated depth of the \oxyseven edge could also be influenced by 
complexity in the spectrum between $\sim 0.7-0.8$ keV,
in particular from Fe inner-shell absorption.

\par\noindent
{\bf Figure 3} \\
\mk MEG observed photon spectrum 
compared to the the best-fitting photoionized absorber model
(red solid line). Also shown is the intrinsic continuum
(blue solid line) modified by Galactic absorption
(neither the data nor model have been corrected for
Galactic absorption). The dashed lines show the expected positions
of some bound-free absorption edges.
The model consists of an
intrinsic continuum which is a broken power law (best-fitting
break energy at 1.04 keV), absorbed by photoionized gas
with best-fitting ionization parameter of $\log{\xi} = 1.76$ (or $\log{U}=0.27$) and
column density $2.06 \times 10^{21} \rm \ cm^{-2}$.
Full details of the model calculations, fitting procedures,
and discussion of the details of the comparison between
data and model can be found in \S\ref{sec:modelling}.
The 0.7--0.9 keV region is very complex so simple
two-edge models fitted to older CCD data (of Seyfert~1s in general)
could have been biased by this complexity.
In particular, note that the apparent emission feature
at $\sim 0.78-0.79 \ {\rm keV}$ (observed) is consistent with unabsorbed continuum
and strong absorption on either side.
There is also an instrumental edge at $\sim 0.8$ keV which may
partly be responsible for the poor fit in this region.
The panel showing the \mgelevenr \resonetwo ($\lambda 9.169\AA$)
line, with data binned at
$0.02\AA$, is shown separately for
clarity, since there are no significant features between 
the end of the previous panel (at 1.15 keV) and the \mgelevenr \resonetwo ($\lambda 9.169\AA$)
absorption line. The remaining
data above $0.5$~keV are binned at $0.08\AA$ and
at $0.04\AA$ below 0.5~keV.
 
\par\noindent
{\bf Figure 4} \\
\chandra MEG photon spectrum for \mk against
wavelength in the source rest-frame (binned at $0.02\AA$,
approximately the MEG FWHM spectral resolution). The corresponding
signal to noise (SNR) ratio per bin is
also shown. The spectrum has {\it not} been corrected for
Galactic absorption.
Many discrete absorption features are identifiable but none are labeled
in this plot since it is primarily for finding the SNR at a particular
wavelength. The same spectrum is shown in Figs. 6 and 7 but this
time with labels corresponding to different arrays of atomic transitions.
See also Figure 3, which shows the $\sim 0.47--1.4$~keV region of the spectrum
with some important atomic features labeled.
 
\par\noindent
{\bf Figure 5} \\
\chandra HEG photon spectrum for \mk against
wavelength in the source rest-frame (binned at $0.02\AA$;
the HEG spectral resolution is $0.012\AA$ FWHM). The corresponding
signal to noise (SNR) ratio per bin is
also shown. The spectrum has {\it not} been corrected for
Galactic absorption. No atomic transitions
are labeled
in this plot since it is primarily for finding the SNR at a particular
wavelength and for confirming features identified in the MEG in
overlapping wavelength regions. Features can be identified using
the labeled MEG spectra in Figs. 6 and  7.

\par\noindent
{\bf Figure 6} \\
\chandra MEG photon spectrum for \mk against
wavelength in the source rest-frame (binned at $0.02\AA$,
approximately the MEG FWHM spectral resolution).
Labels show the Lyman series wavelengths (in blue) for Ar,
S, Si, Mg, Na, Ne,
O, and N. Also shown (in red) are the wavelengths of
the Helium-like triplets (resonance, intercombination and forbidden lines) of
Ar, S, Si, Mg, Ne and O.
 
\par\noindent
{\bf Figure 7} \\
\chandra MEG photon spectrum for \mk against
wavelength in the source rest-frame (binned at $0.02\AA$, 
approximately the MEG FWHM spectral resolution).
Plotted in blue are the wavelengths of He-like
resonance-absorption transitions ($n=1\rightarrow2,3,4,$..) of Ar, S, Si, Mg, 
Na, Ne, O and N. Also shown (in red) are the Balmer series transitions of 
Ar, S and Si.

\par\noindent
{\bf Figure 8} \\
Velocity profiles from combined \chandra MEG and HEG
spectra (except for \oxla which is from MEG data only since it is outside of the
bandpass of the HEG). The profiles are for
some of the strongest absorption features in the data.
It can
be seen that the absorption features are well-described
by the column densities from the
best-fitting photoionization
model (black solid lines), which has $\log{\xi} =1.76$ (or $\log{U}=0.27$) and 
$N_{H}=2.06\times 10^{21} \ {\rm cm ^{-2}}$
(see \S\ref{sec:modelling}). The model profiles were
calculated using a Gaussian with $\sigma = b/\sqrt{2}$, for 
$b=100 \ \rm km \ s^{-1}$, deduced from a curve-of-growth analysis
(see \S\ref{sec:comparison} and \figcog).
A velocity of zero corresponds to
the systemic velocity for \mk (assuming $z=0.0344$).
 A negative velocity here indicates a blueshift relative to systemic.
The centroids of the absorption features lie at around 
$-200 \ {\rm km/s}$ (red), but all the profiles 
extend redward to the systemic velocity
and slightly beyond. FWHM velocities from
Gaussian fitting are given in Table~1 and can be
compared with 
the FWHM MEG velocity resolution
of 352, 496, 550, and 728 km/s for \oxla, \neniner \resonetwo 
($\lambda 13.447\AA$), \nela, and
\mgelevenr \resonetwo ($\lambda 9.169\AA$) respectively.
The velocities of the principal seven kinematic components of the 
UV absorber are also shown (blue) and are
$-422,-328,-259,-62,-22,+34$, and $+124 \ {\rm km \ s^{-1}}$
(see Paper~II).

\par\noindent
{\bf Figure 9} \\
Velocity spectra from combined \chandra MEG and HEG
spectra. The profiles are centered on atomic transitions that 
are not present in the data. 
The solid lines correspond to the best-fitting photoionization
model, which has $\log{\xi} =1.76$ (or $\log{U}=0.27$) and $N_{H}=2.06\times 10^
{21} \ {\rm cm ^{-2}}$
(see \S\ref{sec:modelling}).
Clearly the model does not over-predict
key features which the spectra are centered on (as labeled). 
A velocity of zero corresponds to the 
systemic velocity for \mk (assuming $z=0.0344$). 
A negative velocity here indicates a blueshift 
relative to systemic. The dotted line corresponds to
$-200 \ \rm km \ s^{-1}$, consistent with the
blueshift of absorption lines which {\it are}
detected (\figvelprofone).

\par\noindent
{\bf Figure 10} \\
Curves of growth for various values of the velocity width,
$b$. Plotted for each $b$ is the logarithm of line equivalent width (EW)
per unit wavelength against the logarithm of the product
of ionic column density ($N$), oscillator strength ($f$), and
wavelength ($\lambda$). 
The plotted points are taken from the measured equivalent widths (converted
to Angstroms using the values in Table 1) and predicted ionic column densities
(from the best-fitting XSTAR model described in \S\ref{sec:modelling}). 
These values are consistent with $b= 100 \rm \ km \ s^{-1}$.

\par\noindent
{\bf Figure 11} \\
The observed and interpolated
baseline SED (solid curve) used for photoionization
modeling of Mrk~509. 
Average radio and IR fluxes are from Ward \etal (1987).
The UV data are dereddened fluxes from the
\hst observation (\cite{kraemer02}) and the X-ray data points
are from our simultaneous \chandra \hetg data. The hard X-ray power law has
$\Gamma=1.674$ and has been extended to 500 keV.
The dashed lines are  deviations 
representing modifications to the baseline
SED used to investigate the effects of uncertainties in the
unobserved part of the SED.  The modifications to the
baseline SED correspond to changing the flux at 55 eV
(just above the ionization energy of neutral He) by
$\pm 50\%$.
See \S\ref{sec:sed}
for full details of the construction and applications of the SEDs.
The dotted curve is the 
`mean AGN' SED of  Matthews \& Ferland (1987),
normalized to the same ionizing luminosity (i.e. in 
the range 1--1000 Rydberg) as that of the \mk
baseline SED. 

\par\noindent
{\bf Figure 12} \\
Comparison of MEG data (top panel, raw counts spectrum) binned at $0.16\AA$
with the best-fitting
XSTAR photoionization model (red curve; see \S\ref{sec:modelling})
and the ratio of data to model (bottom panel), showing a good overall
fit to the data. In this representation the various instrumental
edges can be seen, and in particular the poor fit around 2.0--2.5 keV
can be seen to coincide with the large jump in
the X-ray telescope effective area.
 
\par\noindent
{\bf Figure 13} \\
Best-fitting photoionization
model spectrum corresponding to \figxstaroverdata
(see also \figufspeccont).
The ionization parameter is $\log{\xi}=1.76$ (or $\log{U}=0.27$) and $N_{H} = 2.06 \times
10^{21} \rm \ cm^{-2}$. The intrinsic continuum is a
broken power law, and Galactic absorption is included. Full details and other parameters
can be found in the text (\S\ref{sec:modelling}).

\par\noindent
{\bf Figure 14} \\
The $68\%, 90\%$, and $99\%$ confidence contours of
the logarithm of the ionization parameter ($\log{\xi}$) versus
the column density ($N_{H}$) of a photoionized absorber.
The contours are from a joint MEG and HEG spectral fit
to the \mk data, using a broken power-law intrinsic
continuum, a photoionized absorber, and Galactic
absorption. The HEG and MEG spectra were binned at $0.08 \AA$.
See \S\ref{sec:modelling} for details.

\newpage
 
\begin{figure}[h]
\vspace{10pt}
\centerline{\psfig{file=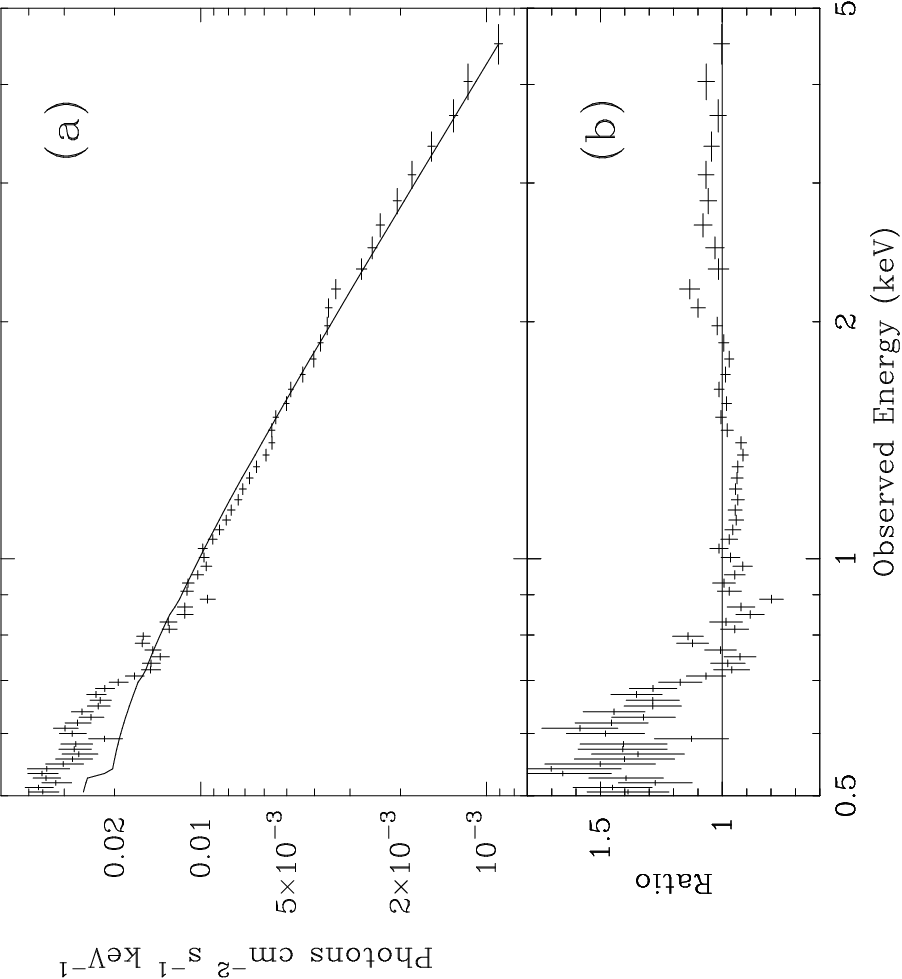,width=8.0in,height=8.0in}
}
\caption{ }
\end{figure}
 
\begin{figure}[h]
\vspace{10pt}
\centerline{\psfig{file=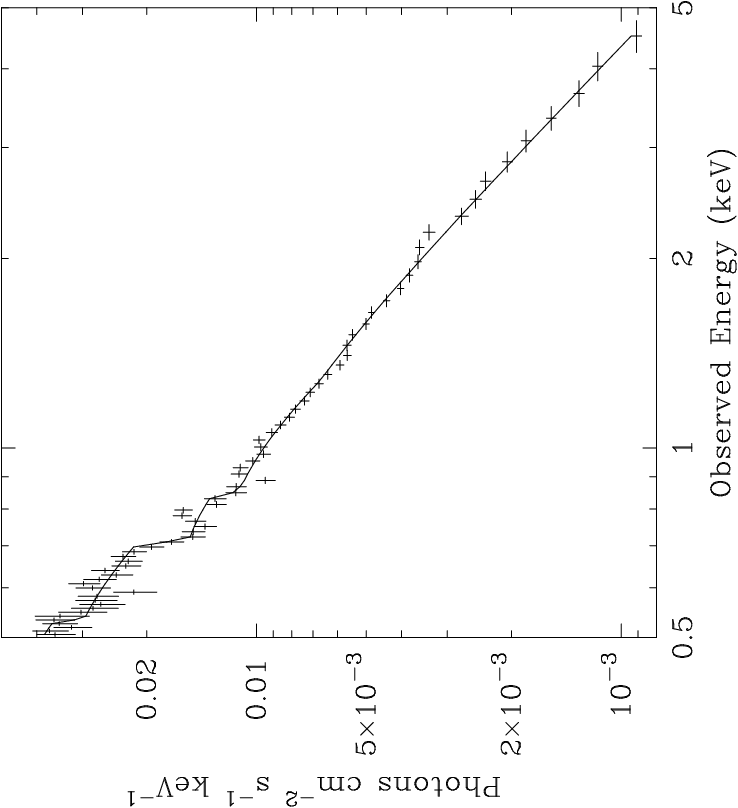,width=8.0in,height=8.0in}
}
\caption{ }
\end{figure}
 
\begin{figure}[h]
\vspace{10pt}
\centerline{\psfig{file=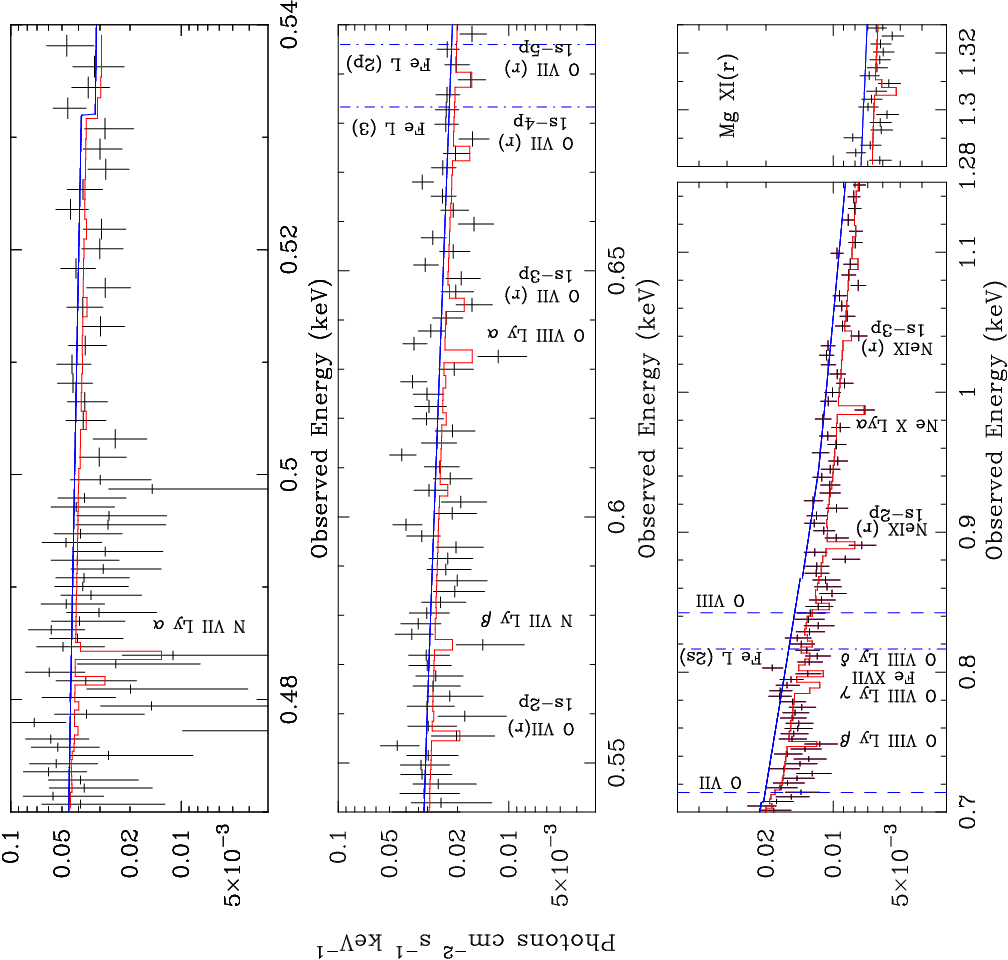,width=8.0in,height=8.0in}
}
\caption{ }
\end{figure}
 
\begin{figure}[h]
\vspace{10pt}
\centerline{\psfig{file=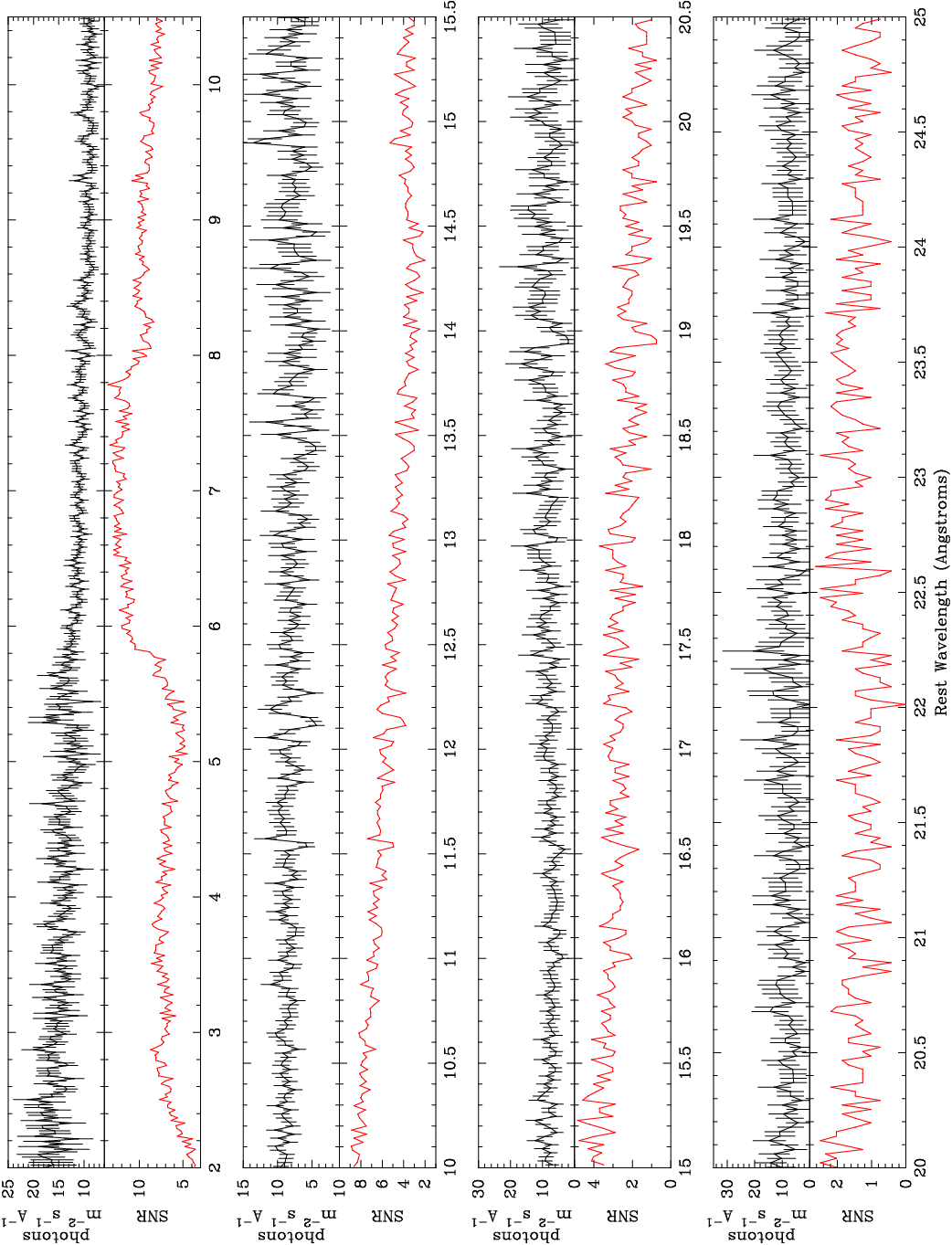,width=8.0in,height=8.0in}
}
\caption{ }
\end{figure}
 
\begin{figure}[h]
\vspace{10pt}
\centerline{\psfig{file=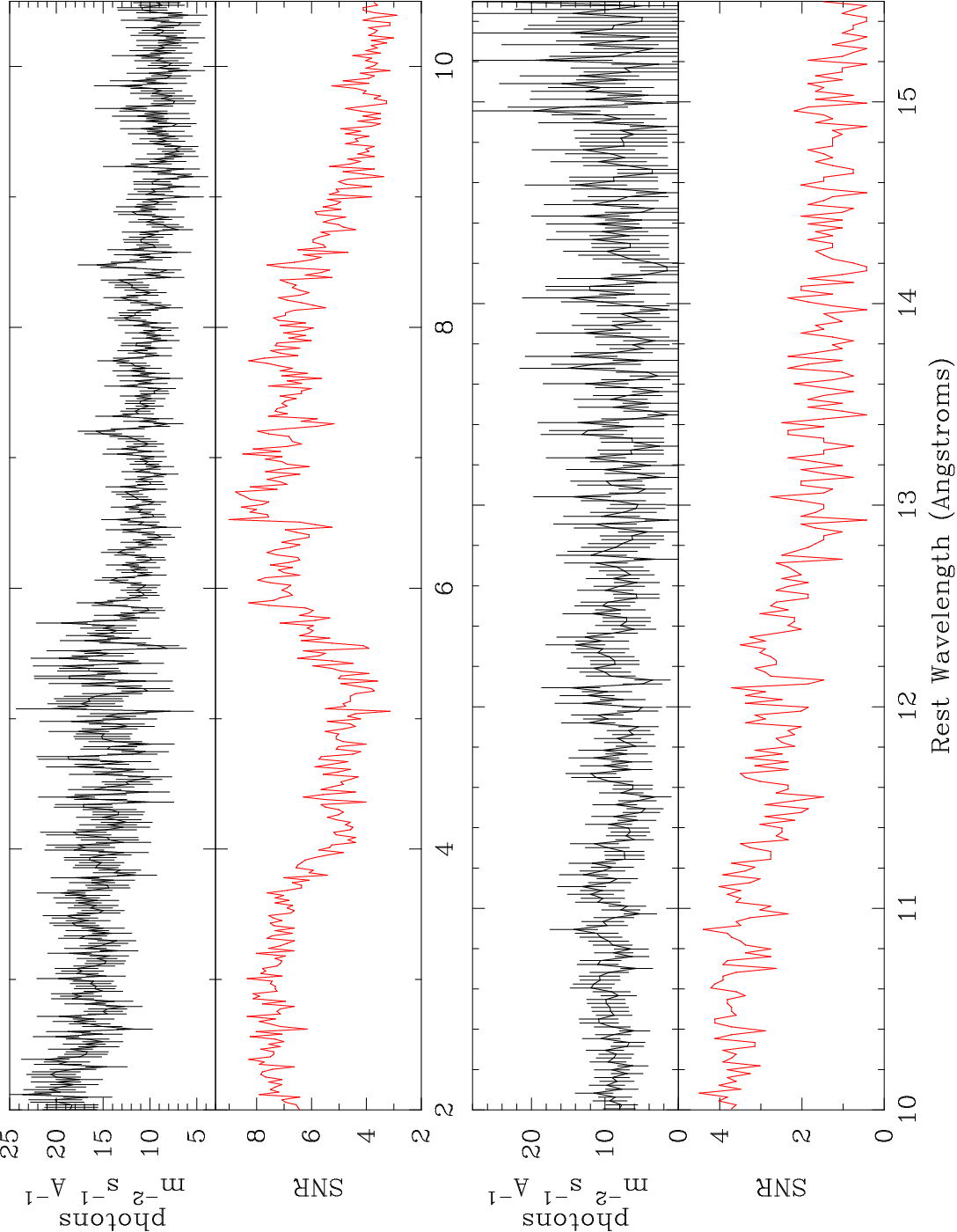,width=8.0in,height=8.0in}
}
\caption{ }
\end{figure}
 
\begin{figure}[h]
\vspace{10pt}
\centerline{\psfig{file=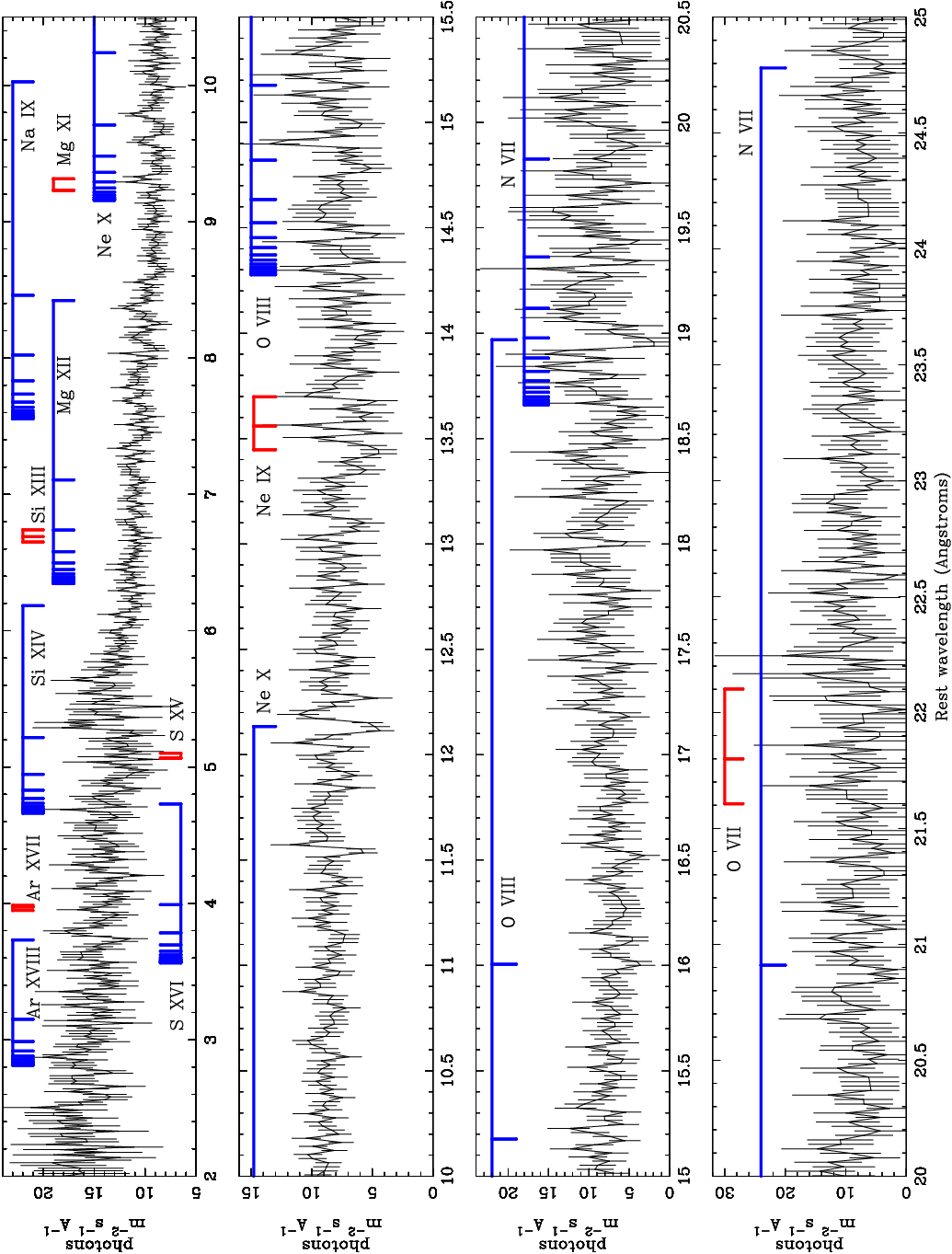,width=8.0in,height=8.0in}
}
\caption{ }
\end{figure}

\begin{figure}[h]
\vspace{10pt}
\centerline{\psfig{file=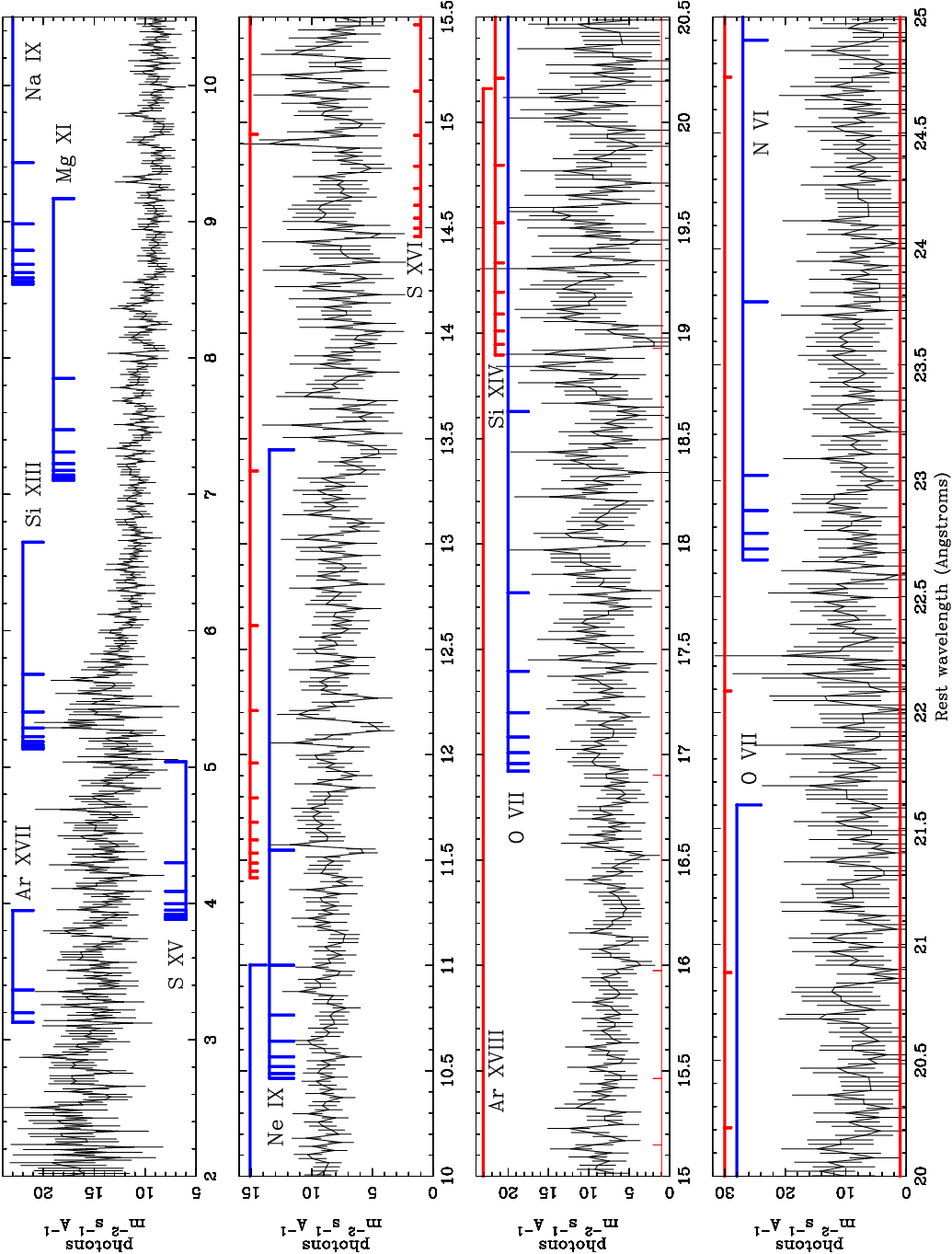,width=8.0in,height=8.0in}
}
\end{figure}

\begin{figure}[h]
\vspace{10pt}
\centerline{\psfig{file=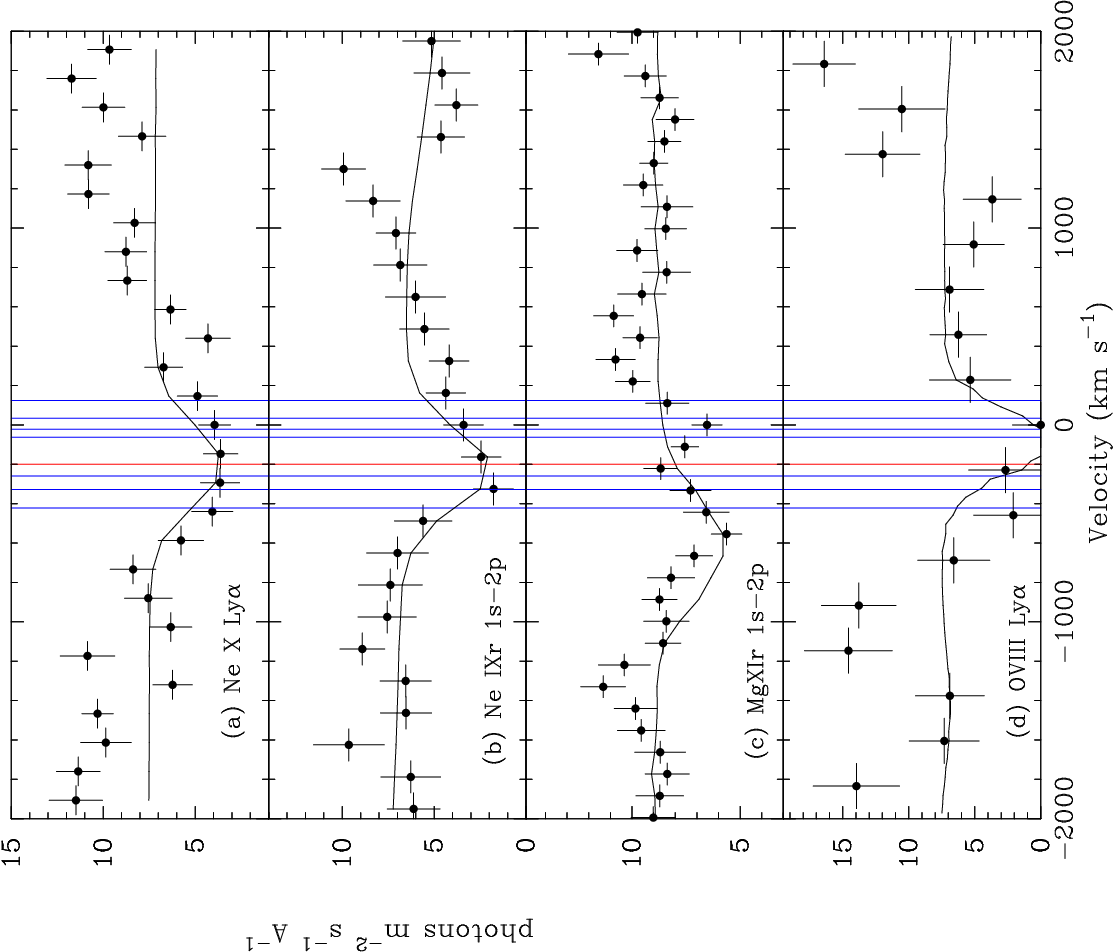,width=8.0in,height=8.0in}
}
\caption{ }
\end{figure}
 
\begin{figure}[h]
\vspace{10pt}
\centerline{\psfig{file=tyf9.ps,width=8.0in,height=8.0in}
}
\caption{ }
\end{figure}
 
\begin{figure}[h]
\vspace{10pt}
\centerline{\psfig{file=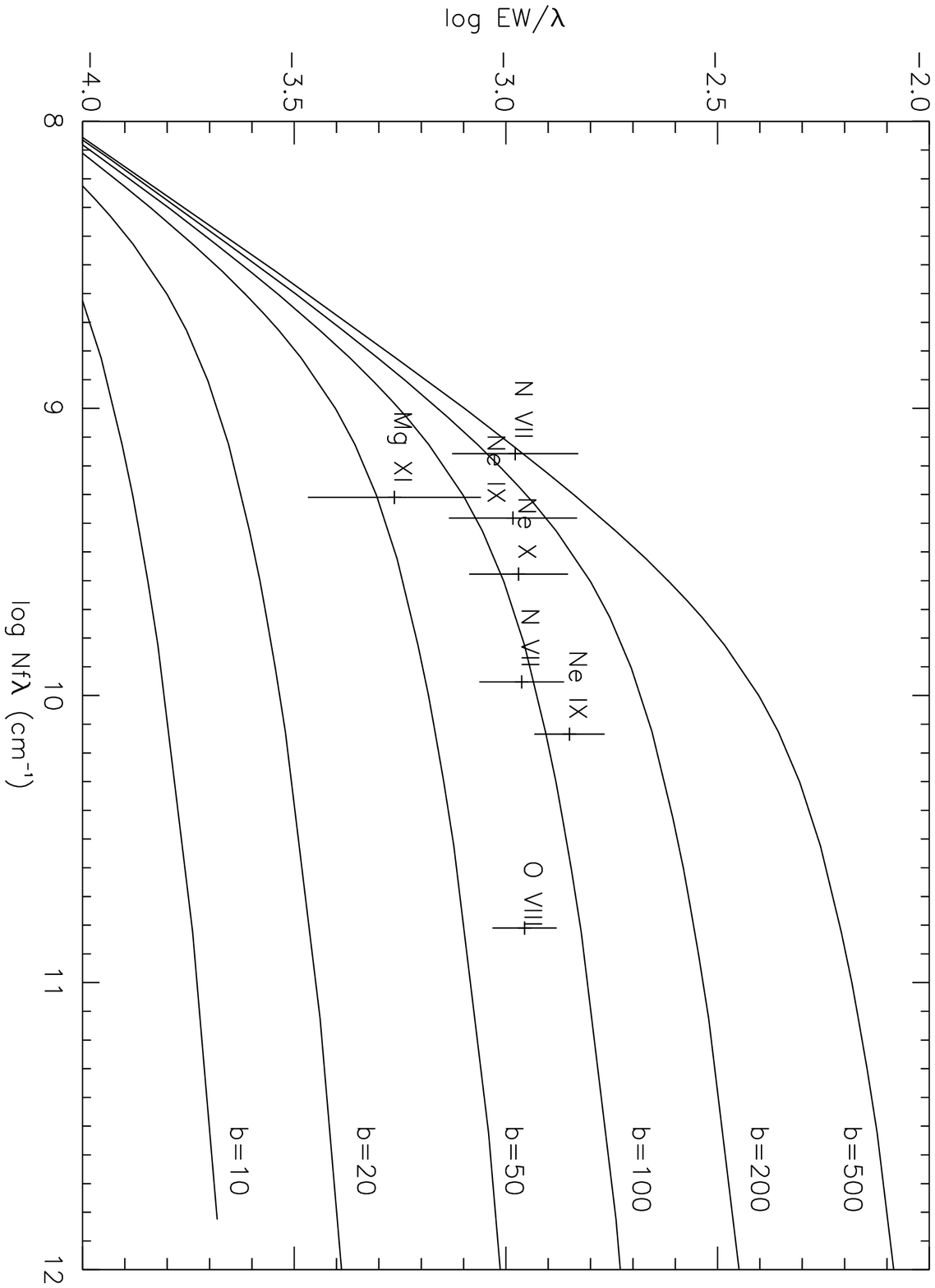,width=8.0in,height=8.0in}
}
\caption{ }
\end{figure}
 
\begin{figure}[h]
\vspace{10pt}
\centerline{\psfig{file=tyf11.ps,width=8.0in,height=8.0in}
}
\caption{ }
\end{figure}
 
\begin{figure}[h]
\vspace{10pt}
\centerline{\psfig{file=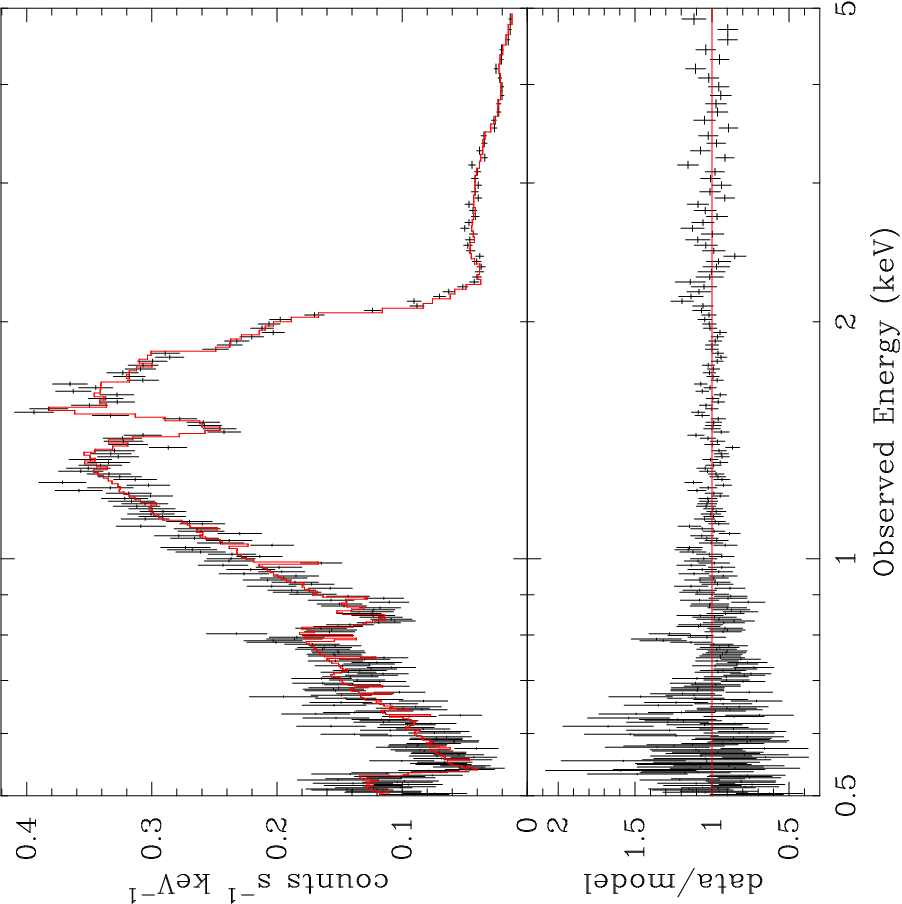,width=8.0in,height=8.0in}
}
\caption{ }
\end{figure}
 
\begin{figure}[h]
\vspace{10pt}
\centerline{\psfig{file=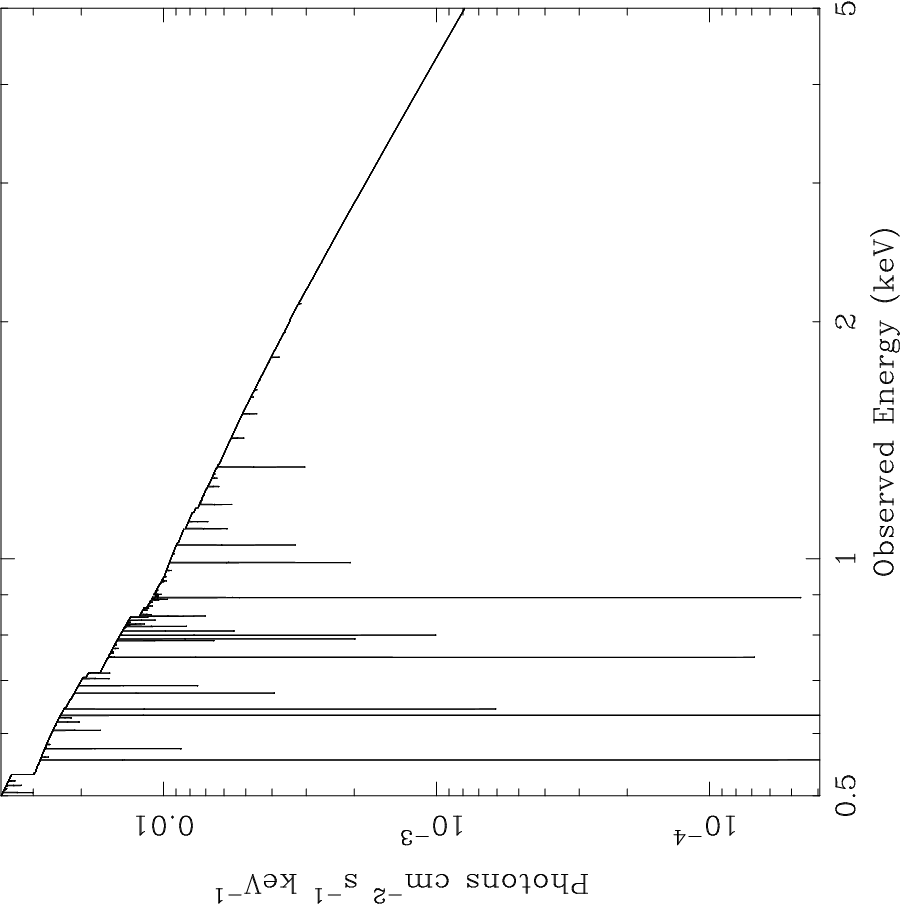,width=8.0in,height=8.0in}
}
\caption{ }
\end{figure}

\begin{figure}[h]
\vspace{10pt}
\centerline{\psfig{file=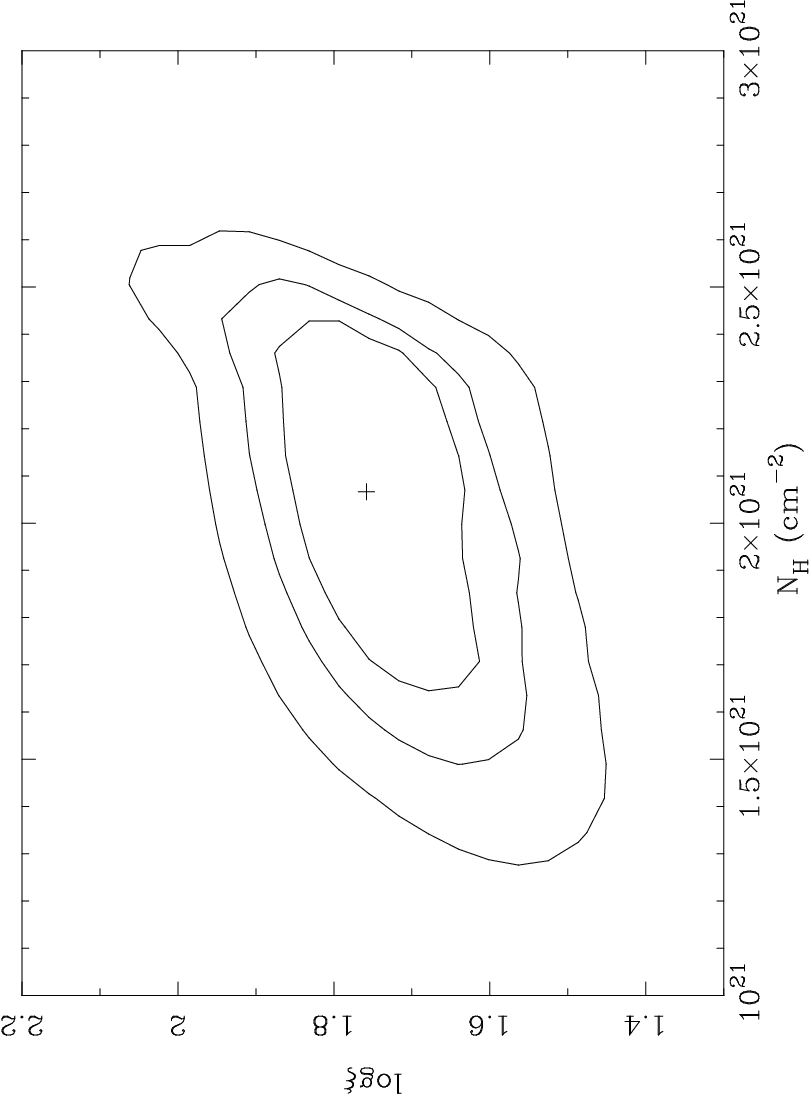,width=8.0in,height=8.0in}
}
\caption{ }
\end{figure}

\end{document}